\documentclass[conference]{IEEEtran}
\usepackage[T1]{fontenc}
\usepackage[utf8]{inputenc}
\usepackage{letltxmacro}
\usepackage{todonotes}
\usepackage{xspace}
\usepackage{hyperref}
\usepackage{enumerate}
\usepackage{caption}% http://ctan.org/pkg/caption
\usepackage{graphicx}
\usepackage{xcolor,colortbl}
\usepackage{pifont}
\usepackage{amssymb} %for math symbols
\usepackage[update,prepend]{epstopdf}
\usepackage{listings}
\usepackage{array}
\usepackage[ruled,vlined,linesnumbered]{algorithm2e}
\SetAlgoSkip{}
\graphicspath{{sections/figures/}}
\definecolor{mGreen}{rgb}{0,0.6,0}
\definecolor{mGray}{rgb}{0.5,0.5,0.5}
\definecolor{mPurple}{rgb}{0.58,0,0.82}
\definecolor{backgroundColour}{rgb}{0.95,0.95,0.92}
\definecolor{notecolor}{rgb}{0,0.0,0} 
\lstdefinestyle{CStyle}{
	backgroundcolor=\color{backgroundColour},   
	commentstyle=\color{mGreen},
	keywordstyle=\color{magenta},
	numberstyle=\tiny\color{mGray},
	stringstyle=\color{mPurple},
	basicstyle=\footnotesize\ttfamily,
	breakatwhitespace=false, 
	breaklines=true,                 
	captionpos=b,                    
	keepspaces=true,                 
	numbers=left,                    
	numbersep=3pt,                  
	showspaces=false,                
	showstringspaces=false,
	showtabs=false,                  
	tabsize=2,
	aboveskip=2mm,
	belowskip=2mm,
	language=C,
	xleftmargin=0.2in,
	xrightmargin=0.2in
}

\newcommand{\xmark}{\ding{55}}
\newcommand{\cmark}{\ding{51}}
\newif\ifabridged
\newif\ifnotabridged
\newif\ifanonymous
\newif\ifnotanonymous

\ifdefined\isabridged
\abridgedtrue
\fi

\ifabridged\notabridgedfalse
\else\notabridgedtrue
\fi

\ifdefined\isanonymous
\anonymoustrue
\fi

\ifanonymous\notanonymousfalse
\else\notanonymoustrue
\fi

\LetLtxMacro{\todonote}{\todo}
\renewcommand{\todo}[2][]
{\todonote[inline, caption={#2}, size=\footnotesize, #1]
{\renewcommand{\baselinestretch}{0.5}\selectfont#2\par}}

\newcommand{\etal}{\emph{et al.}}
\newcommand{\eg}{\emph{e.g.}}
\newcommand{\ie}{\emph{i.e.}}

\title{Exploitation Techniques and Defenses \\for Data-Oriented Attacks}%Exploitation Techniques and Defenses for Data-Oriented Attacks

\author{
	\IEEEauthorblockN{Long Cheng\IEEEauthorrefmark{1}, Hans Liljestrand\IEEEauthorrefmark{2}, Thomas Nyman\IEEEauthorrefmark{2}, Yu Tsung Lee\IEEEauthorrefmark{4}, \\Danfeng (Daphne) Yao\IEEEauthorrefmark{3}, Trent Jaeger\IEEEauthorrefmark{4} and N. Asokan\IEEEauthorrefmark{3}}
	
	\IEEEauthorblockA{\IEEEauthorrefmark{1}School of Computing, Clemson University, USA}  
	\IEEEauthorblockA{\IEEEauthorrefmark{2}Department of Computer Science, Aalto University, Finland}
	\IEEEauthorblockA{\IEEEauthorrefmark{3}Department of Computer Science, Virginia Tech, USA}
	\IEEEauthorblockA{\IEEEauthorrefmark{4}Department of Computer Science and Engineering, Pennsylvania State University, USA}
	
}

\ifnotanonymous
\fi

\begin{document}
\maketitle

% Abstract
\begin{abstract}
 
Data-oriented attacks manipulate non-control data to alter a program's benign behavior without violating its control-flow integrity. It has been shown that such attacks can cause significant damage even in the presence of control-flow defense mechanisms.  However, these threats have not been adequately addressed. In this SoK paper, we first map data-oriented exploits, including Data-Oriented Programming (DOP) attacks, to their assumptions/requirements and attack capabilities. We also compare known defenses against these attacks, in terms of approach, detection capabilities, overhead, and compatibility. Then, we experimentally assess the feasibility of a detection approach that is based on the Intel Processor Trace (PT) technology. PT only traces control flows, thus, is generally believed to be not useful for data-oriented security. However, our work reveals that data-oriented attacks (in particular the recent DOP attacks) may generate side-effects on control-flow behavior in multiple dimensions, which manifest in PT traces. Based on this evaluation, we discuss challenges for building deployable data-oriented defenses and open research questions.

\end{abstract}

% Introduction
\section{Introduction}

%Buffer overflows are notorious 
Memory-corruption vulnerabilities are one of the most common attack vectors used to compromise computer systems. Such vulnerabilities can be exploited in different ways, which potentially allow attackers to perform arbitrary code execution and data manipulation. 
%Memory-corruption vulnerabilities would remain an unsolved problem for a long time. This is because: 1) memory-unsafe languages such as C/C++ are still widely used today and remain common sources of security vulnerabilities; and 2) despite considerable research in past years, existing mechanisms suffer one or more issues including large performance overhead, inaccurate/incomplete coverage, or imperfect compatibility~\cite{Szekeres:2013:SEW, Sinha:2018:ISCA}. 
Existing memory corruption attacks can be broadly classified into two categories: i) control-flow attacks~\cite{Francillon:2008:CIA, Shacham:2007:GIF, roemer:tissec12} and ii) data-oriented attacks (also known as non-control data attacks)~\cite{Szekeres:2013:SEW, Chen:2005:NAR, Hu:2015:DOA, Hu16, Morton:EuroSP:2018}. 
Both types of attacks can cause significant damages to a victim system~\cite{Carlini:2015:CBE}.

Control-flow attacks corrupt control data (\eg, return address or code pointer) in a program's memory space to divert the program's control flow, including
malicious code injection~\cite{Francillon:2008:CIA}, code reuse~\cite{Shacham:2007:GIF}, and Return-Oriented Programming (ROP)~\cite{roemer:tissec12}.
To counter these attacks, many defense mechanisms have been proposed, such as stack canaries~\cite{Cowan:1998:SAA}, Data Execution Prevention (DEP)~\cite{DEP:Microsoft}, Address Space Layout Randomization (ASLR)~\cite{Shacham:2004:EAR},  Control-Flow Integrity (CFI)~\cite{Abadi:2005:CI}, Return-Flow Guard (RFG)~\cite{RFG:Microsoft}, Intel's CET~\cite{Intel:CET} and MPX~\cite{MPX}. In particular, CFI-based solutions~\cite{Burow:2017:CIP} have received considerable attention in the last decade. The idea is to ensure that the runtime program execution always follows 
a valid path in the program's Control-Flow Graph (CFG), by enforcing security policies on indirect control transfer instructions \textcolor{notecolor}{(\eg, ret/jmp)}. 

%Carlini~\etal~\cite{Carlini:2015:CBE} have recently revealed fundamental limits on the effectiveness of CFI, and presented the Control-Flow Bending (CFB) which allows an attacker to "bend" the control-flow of a program but adheres to CFI's security policies, \ie, modifying indirect branch targets that are valid as defined by a CFI policy. (a.k.a.  non-control data attacks

In contrast to control-flow attacks, data-oriented attacks~\cite{Chen:2005:NAR} change a program's benign behavior by manipulating the program's non-control data (\eg, a data variable/pointer which does not contain the target address for a control transfer) without violating its control-flow integrity. The attack objectives include: 1) information disclosure (\eg, leaking passwords or private keys); 2) privilege escalation (\eg, by manipulating user identity data)~\cite{Chen:2005:NAR};  3) performance degradation (\eg, resource wastage attack)~\cite{Baliga:2007:SP}; and 4) bypassing security mitigation mechanisms~\cite{Xiao:SecureComm:2015}.

As launching control-flow attacks becomes increasingly difficult due to many deployed defenses against control-flow hijacking, data-oriented attacks are likely to become an appealing attack technique for system compromise~\cite{Schlesinger11, Hu:2015:DOA, Xiao:SecureComm:2015, Hu16, DBLP:NymanDZLPAS17, Morton:EuroSP:2018}.
Data-oriented attacks can be as simple as flipping a bit of a variable. However, they can be equally powerful and effective as control-flow attacks~\cite{song:kenali:ndss}. For example, arbitrary code-execution attacks are possible if an attacker could corrupt parameters of system calls (\eg, \verb1execve()1)~\cite{Carlini:2015:CBE}.  
Recently, Hu~\etal~\cite{Hu16} proposed Data-Oriented Programming (DOP), a systematic technique to construct expressive (\ie, Turing-complete) non-control data exploits. 
Ispoglou~\etal~\cite{Ispoglou:2018:BOP} presented the Block-Oriented Programming (BOP), a code reuse technique that utilizes basic blocks
as gadgets along valid execution paths in the target binary to generate data-oriented exploits.
Though data-oriented attacks have been known for a long time, the threats posed by them have not been adequately addressed due to the fact that most previous defense mechanisms focus on preventing control-flow exploits. 

%Data-Flow Integrity (DFI)~\cite{Castro06} applies the static analysis to compute the data-flow graph of a program, and aims to protect the program against data-oriented attacks by enforcing runtime data-flow policies. \textcolor{notecolor}{However, software based DFI incurs a high performance overhead. In addition, DFI usually overestimates the data-flow policies and thus cannot completely prevent data manipulation.}

The motivation of this paper is to systematize the current knowledge about exploitation techniques of data-oriented attacks and the current applicable defense mechanisms. Unlike prior systematization of knowledge (SoK) papers~\cite{Szekeres:2013:SEW, Larsen14, SoK:2018} related to memory corruption vulnerabilities, our work specifically focuses on data-oriented attacks. 
In addition to generic memory corruption prevention mechanisms discussed in~\cite{Szekeres:2013:SEW, Larsen14, SoK:2018} such as memory safety, software compartmentalization, and address/code space randomization, we mainly discuss recently proposed defenses against data-oriented attacks. In particular, we experimentally assess the possibility of a detection approach based on hardware-assisted control-flow tracing (\eg, Intel PT) against advanced data-oriented attacks.  
 
Our technical contributions are as follows.
\begin{enumerate}[*]
\item We systematize the current knowledge about data-oriented exploitation techniques with a focus on the recent DOP attacks. We demystify the DOP exploitation technique by using the ProFTPd DOP attack~\cite{Hu:2015:DOA} as a case study, and provide an intuitive and detailed explanation of this attack by analyzing its constituent steps. We also discuss representative data-oriented exploits including their assumptions/requirements and attack capabilities (Section~\ref{sec2}). 

\item We present a three-stage model for data-oriented attacks and discuss defense techniques according to different stages. Then, we provide a comparative analysis of recent defensive approaches specially focusing on data-oriented attacks (Section~\ref{defenseoverview}).
%.  and specially focusing on data-oriented attacks

\item We investigate the possibility of using the hardware-assisted control-flow tracing for program anomaly detection against data-oriented attacks, and study the limitation of protection offered by such approach (Section~\ref{feasibility-overview}).
	% Our study stimulates a broader discussion about advanced defenses against data-oriented attacks. present preliminary results using program anomaly detection to detect anomalous program behaviors in DOP attacks.
\item We experimentally assess the feasibility of a detection approach based on control-flow tracing in particular against DOP attacks. We demonstrate that DOP exploits may generate side-effects on control-flow behaviors in multiple dimensions, and thus are more likely to be detected by a program anomaly detector (Section~\ref{sec5:detection}). 
%We show that a well-designed program anomaly detection system increases the difficulties that a data-oriented attack bypasses the detection system.
We also discuss some open research problems and unsolved challenges (Section~\ref{sec7:discussion}). 
\end{enumerate}

% Data-oriented Attacks
\section{Data-oriented Attacks}\label{sec2}

In this section, we introduce two categories of exploitation techniques to launch data-oriented attacks (Section~\ref{sec2:1}). We reproduce a real-world DOP attack against the ProFTPD FTP-server~\cite{Hu16} and present a detailed description of the attack to demonstrate how the complex attack achieves rich expressiveness (Section~\ref{sec2:proftpd}). Then, we map representative data-oriented exploits in the literature to their assumptions/requirements and attack capabilities (Section~\ref{sec2:3}).

\subsection{Classification of data-oriented attacks}\label{sec2:1}

We classify data-oriented attacks\footnote{\scriptsize In this work, we mainly focus our investigation on data-oriented attacks that are caused by memory-corruption vulnerabilities. Data-only attacks that are caused by hardware transient faults or logic errors in code are beyond the scope of this work.} into two categories based on how attackers manipulate the non-control data in memory space: 1) Direct Data Manipulation (DDM); and 2) Data-Oriented Programming (DOP). 

\emph{1) DDM} refers to a category of attacks in which an attacker directly manipulates the target data to accomplish the malicious goal. It requires the attacker to know the precise memory address of the target non-control-data. The address \textcolor{notecolor}{or offset to a known location utilized in the attack} can be derived directly from binary analysis (\eg, global variable with a deterministic address) or by reusing the runtime randomized address stored in memory~\cite{Hu:2015:DOA}.
Several types of memory corruption vulnerabilities, \eg, format string vulnerabilities, buffer overflows, integer overflows, and double free vulnerabilities~\cite{SoK:2018}, allow attackers to directly overwrite memory locations within the address space of a vulnerable application. Chen~\etal~\cite{Chen:2005:NAR} revealed that DDM attacks can corrupt a variety of security-critical variables including user identity data, configuration data, user input data, and decision-making data, which change the program's benign behavior or cause the program to inadvertently leak sensitive data.

%\vspace{-11pt}

\lstdefinestyle{CStyleListing1}{
	backgroundcolor=\color{backgroundColour},   
	commentstyle=\color{mGreen},
	keywordstyle=\color{magenta},
	numberstyle=\tiny\color{mGray},
	stringstyle=\color{mPurple},
	basicstyle=\footnotesize\ttfamily,
	breakatwhitespace=false, 
	breaklines=true,                 
	captionpos=b,                    
	keepspaces=true,                 
	numbers=left,                    
	numbersep=3pt,                  
	showspaces=false,                
	showstringspaces=false,
	showtabs=false,                  
	tabsize=2,
	aboveskip=2mm,
	belowskip=2mm,
	language=C,
	xleftmargin=0.1in,
	xrightmargin=0.1in
}

\begin{lstlisting}[caption={DDM attack in a vulnerable SSH server~\cite{Chen:2005:NAR}},label={lst:authen}, style=CStyleListing1, basicstyle=\scriptsize] 
void do_authentication(char *user, ...) {
	...
	int authenticated = 0;
	...
	while(!authenticated){
		type = packet_read();//Corrupt authenticated
		/*Calls detect_attack() internally*/
		switch(type){
			...
			case SSH_CMSG_AUTH_PASSWORD:
				if(auth_password(user, password)){
					authenticated = 1;
					break;}
			case ...
		}
		if(authenticated) break;
	}
	do_authenticated(pw); 
	/*Perform session preparation*/
}
\end{lstlisting}
%\vspace{-11pt}

Listing~\ref{lst:authen} illustrates an example of the attack on decision-making data in SSH server, which was first reported in~\cite{Chen:2005:NAR}. A local flag variable \texttt{authenticated} is used to indicate whether a remote user has passed the authentication (line 3). 
An integer overflow vulnerability exists in the \texttt{detect\_attack()} function, which is internally invoked whenever the \texttt{packet\_read()} function is called (line 6). When the vulnerable function is invoked, an attacker is able to corrupt the \texttt{authenticated} variable to a non-zero value, which bypasses the user authentication (line 16).

\emph{2) DOP} is an advanced technique to construct expressive non-control data exploits~\cite{Hu16}. It allows an attacker to perform arbitrary computations in program memory by chaining the execution of short sequences of instructions (referred to as \emph{data-oriented or DOP gadgets}). The idea is to reuse the code in these gadgets for malicious purposes other than the developer's original intent. Similarly, Block-Oriented Programming (BOP)~\cite{Ispoglou:2018:BOP} constructs exploit programs by chaining BOP gadgets without violating CFI, where each BOP gadget corresponds to a basic block that contains a DOP gadget. Without loss of generality, we use DOP to represent this exploitation technique, which misinterprets multiple gadgets and chains these gadgets together by one or more dispatchers to achieve the desired outcome. 

Typically, a DOP attack corrupts several memory locations in a program and involves multiple steps. To understand the complexity and the expressiveness of the DOP technique, we dissect a real-world DOP attack in Section~\ref{sec2:proftpd}.

%\subsection{Multi-step DDM, DOP/BOP, and BOPC}
\label{sec:multistep-DDM}
%The adversary may do this repeatedly at the same instruction.
There also exists multi-step DDM attacks, where an adversary exploits memory corruption vulnerabilities multiple times to write data to adversary-chosen memory locations. For example, suppose an attacker needs to change two decision-making variables while the vulnerability only allows the attacker to change one value each time. It requires a 2-step DDM. Morton~\etal~\cite{Morton:EuroSP:2018}~recently demonstrated a multi-step DDM with Nginx (listed in Table~\ref{RepresentativeAttacks}). The attack leverages memory errors to modify global configuration data structures in web servers. Constructing a faux SSL Config struct in Nginx requires as many as 16 connections (\ie, 16-step DDM)~\cite{Morton:EuroSP:2018}.

Like the DOP attack, a multi-step DDM attack violates data-flow integrity. DDM is a pre-requisite for DOP. However, DOP is much more complex than the multi-step DDM. We summarize their key differences in the following. 
%Our comparison also includes the new BOP technique by Ispoglou~\etal~\cite{Ispoglou:2018:BOP}. 

\begin{enumerate}[*]
	\item
	{\bf Gadgets and code reuse.}
	DOP/BOP attacks involve reusing code execution through CFI-compatible gadgets.
	Multi-step DDM hinges on direct memory writes and does not involve any gadget executions. 
	%In multi-step DDMs, the DFI violation is limited to the locations of vulnerable instructions, whereas DOP's DFI violation spreads across the entire gadget chain.
	
	% The DOP paper classifies different types of short code sequences based on there semantic operation, e.g., addition, dereference, conditional. For example, “server->total += *size” is originally used to keep the total size of received data. Under DOP attack, attacker uses this statement to change value of a structure (by letting the server points to the target structure, and manipulate the value of *size). On the contrary, multistep DDM does not misinterpret existing code sequences for malicious purposes, they directly change target data multiple times.  
	\item
	{\bf Stitching mechanism and ordering constraint.}
	In DOP and BOP attacks, how to orderly stitch gadgets to form a meaningful attack is important. Multi-step DDM attacks, \eg, crafting and sending multiple attack payloads to manipulate memory values, do not need any special stitching mechanism (and thus there is no ordering constraint).  
	
	%\item
	%{\bf Following pointers.} 
	%Multi-step DDM assumes the adversary does not have the ability to follow pointers in the data structure. In contrast, the ability to manipulate pointer content and following pointers is crucial for DOP and BOP attacks, where an adversary uses the initial vulnerability to setup pointers to perform a sequence of write operations.

\end{enumerate}

A significant contribution by Ispoglou~\etal~in~\cite{Ispoglou:2018:BOP} is the block-oriented programming compiler (BOPC).
BOPC is the first compiler technique that automates the BOP/DOP attack generation (given the arbitrary memory write vulnerability).
With the automatically generated attack payloads by the compiler, an attacker first performs a series of DDMs to modify memory and then launches a BOP/DOP attack by chaining gadgets that leverage memory manipulation via DDMs.

%BOPC

\subsection{Demystifying the ProFTPd DOP attack}\label{sec2:proftpd}
%the purpose of this example is to give an instance of 3 stages and lead to discussion of DOP requirements, the concept of gadgets.

We use the ProFTPd DOP attack crafted by Hu~\etal~\cite{Hu16} to illustrate the typical flow of DOP attacks. 
%It was crafted by  to demonstrate how a DOP attack can defeat randomization defenses (such as ASLR) without leaking any address in a real-world program ProFTPD. 
The goal of this DOP attack is to bypass randomization defenses (such as ASLR~\cite{Shacham:2004:EAR}), and then leak the server's OpenSSL private key.
The private key is stored on the heap with an unpredictable layout, which hinders the attacker from reading out the private key from the heap directly. 
Though the key is stored in a randomized memory region, it can be accessed via a chain of 8 pointers. As long as the base pointer is not randomized, \eg, when the position independent executables (PIE) feature is disabled, it is possible to exfiltrate the private key by starting from the OpenSSL context base pointer (\ie, a known location of the static variable \texttt{ssl\_ctx}) and recursively de-referencing 7 times within the server's memory space. 

%in case that the program is not compiled as Position Independent Executables (PIE) 
%which does not randomize the full program image
%(\eg, PIE is disabled by default on 32bit OS to prevent a 10\% performance degradation~\cite{Szekeres:2013:SEW})}. 

\begin{figure*}[!h] %\vspace{-10pt}
	\centering
	\includegraphics[width=0.78\textwidth]{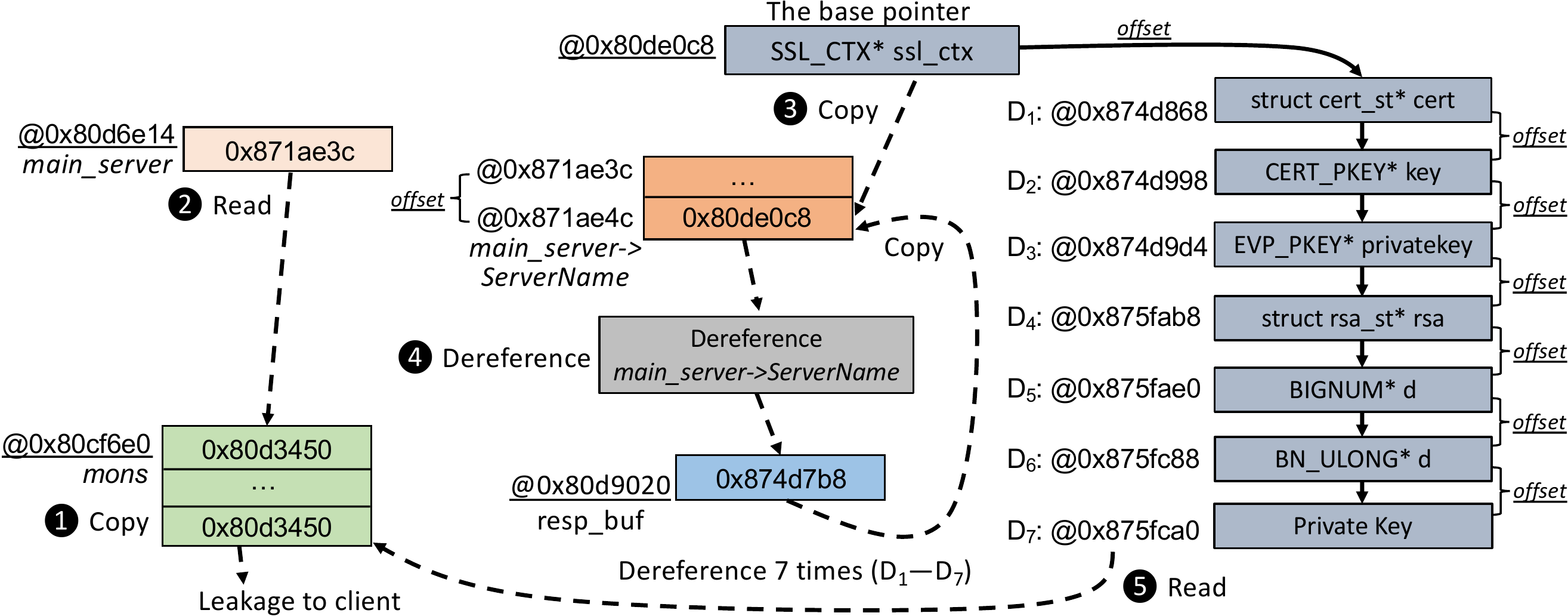} \vspace{-1pt}
	\caption{ProFTPd DOP attack flow. An attacker needs to know the underlined addresses and offsets to launch the attack.}  
	\label{fig:proftpd} \vspace{-15pt}
\end{figure*}

\subsubsection{ProFTPd vulnerability}
ProFTPD versions 1.2 and 1.3 have a stack-based buffer overflow vulnerability in the \texttt{sreplace} function (CVE-2006-5815~\cite{ProFTPD}).
The overflow can be exploited by an attacker to obtain an arbitrary write primitive. The server program provides a feature to display customized messages when a user enters a directory. The message content is saved in \texttt{.message} file in each directory.
It can be edited by any user with write-access to the directory. The \texttt{.message} file can contain special characters (\ie, specifiers) which will be replaced with dynamic content such as time/date and server name by the \texttt{sreplace} function. For example, the string "\%V" in \texttt{.message} will be replaced by \texttt{main\_server->ServerName}, and "\%T" will be replaced by the current time and date.
Changing the working directory with a \texttt{CWD} command triggers the processing of \texttt{.message} file, and subsequently triggers the invocation of the \texttt{sreplace} function. To trigger a memory error in the vulnerable function, the attacker crafts attack payloads to modify the content of the \texttt{.message} file, and then send \texttt{CWD} commands to the server.

\begin{lstlisting}[caption={The vulnerable function in ProFTPd},label={lst:sreplace},style=CStyleListing1,basicstyle=\scriptsize] 
char *sstrncpy(char *dest, const char *src, size_t n) {
	register char *d = dest;
	for (; *src && n > 1; n--)
		*d++ = *src++;
	...
}
char *sreplace(char *s, ...) {
	...
	char *m,*r,*src = s,*cp;
	char **mptr,**rptr;
	char *marr[33],*rarr[33];
	char buf[BUF_MAX] = {'\0'}, *pbuf = NULL;
	size_t mlen=0, rlen=0, blen; cp=buf;
	...
	while(*src){
		for(mptr=marr, rptr=rarr; *mptr; mptr++, rptr++) {
			mlen = strlen(*mptr);
			rlen = strlen(*rptr);
			if(strncmp(src,*mptr,mlen)==0){//check specifiers
				sstrncpy(cp,*rptr,blen-strlen(pbuf)); //replace a specifier with dynamic content stored in *rptr 
				if(((cp + rlen) - pbuf + 1) > blen){   
					cp = pbuf + blen - 1; ...
				} /*Overflow Check*/ 
				...
				src += mlen;
				break;
			}
		}
		if(!*mptr) {
			if((cp - pbuf + 1) > blen){ //off-by-one error
				cp = pbuf + blen - 1; ...
			} /*Overflow Check*/
			*cp++ = *src++; 
		}
	}
}
\end{lstlisting}

Listing~\ref{lst:sreplace} shows the vulnerable \texttt{sreplace} function. The vulnerability is introduced by an off-by-one comparison bug in line 30, which allows attackers to modify the program memory. A defective overflow check in lines 29-34 is performed to detect any attempt to write outside the buffer boundary. When writing to the last character of the buffer \texttt{buf}, \texttt{(cp-pbuf+1)} equals to \texttt{blen}. Thus, the predicate in line 30 returns \texttt{false}, and the string terminator is overwritten in line 33. Consequently, the string is not properly terminated inside the buffer because the buffer's last character has been overwritten with a non-zero byte. In the next iteration of the \texttt{while} loop, the input \texttt{blen-strlen(pbuf)} of the \texttt{sstrncpy} function becomes negative, which will be interpreted as a large unsigned integer (in line 20). Hence, the invocation of \texttt{sstrncpy} overflows outside buffer bounds into the stack and overwrites local variables such as \texttt{cp}. 
Both the source (\ie, \texttt{*rptr}) and the destination (\ie, \texttt{cp}) of the string copy function, \ie, \texttt{sstrncpy} in line 20, are under the control of the attacker, where \texttt{*rptr} can be manipulated by the attacker through specifying special characters in \texttt{.message} (\eg, "\%C" will be replaced by an attacker-specified directory name).
As a result, the vulnerability allows the attacker to control the source, destination, and number of bytes copied by subsequent iterations of the \texttt{while} loop in lines 15-35.

%empowers the attacker with arbitrary memory writing. 

%in the correct order, one at a time to achieve the desired outcome

\subsubsection{The attack flow}
%%match the 3 stages 

Fig.~\ref{fig:proftpd} shows a step-by-step description of the ProFTPd DOP attack, where the underlined addresses and offsets are required to be known before launching the attack.
 
The attacker interacts with the server program (over the course of numerous FTP commands) to corrupt program memory by repeatedly exploiting the buffer overflow vulnerability. In this scenario, the command handler \texttt{cmd\_loop} in ProFTPd serves as the data-oriented gadget dispatcher. In each iteration, the attacker triggers the execution of targeted gadgets by sending a crafted attack payload to the server program, \eg, the dereference gadget \texttt{*d++=*src++} located in \texttt{sstrncpy} (line 4 in Listing~\ref{lst:sreplace}).
We reproduced the ProFTPd DOP attack, and observed that the vulnerable function \texttt{sreplace} is called more than 180 times during the attack.

% to change the flow chart to 6 steps, combine 2 and 4.

Over the course of the attack, based on the known addresses and offsets (underlined in Fig.~\ref{fig:proftpd}) extracted from the ProFTPd binary, the attacker systematically corrupts program memory to construct a DOP program out of individual operations. The main steps, shown in Fig.~\ref{fig:proftpd}, are described as follows. 
%The interactive attack includes the following main steps.
\begin{enumerate} [\hspace{-12pt}(P1)]
	\item[] \ding{182} To read data from arbitrary addresses in the server, the attacker needs to overwrite string pointers used by a public output function (\eg, \texttt{send}). To this end, the attacker manipulates 12 pointers in a local static \texttt{mons} array located at \texttt{0x80cf6e0} to a global writable location (\ie, \textcolor{notecolor}{the attacker specifies this location, denoted by \texttt{G\_PTR}}). As shown in Fig.~\ref{fig:proftpd}, the \texttt{mons} array is filled with \texttt{G\_PTR}'s address \texttt{0x80d3450}.
	Thus, when the server returns the date information to the client, it prints the value pointed by \texttt{G\_PTR}. This step builds an exfiltration channel which can leak information from the server to the network.
	%The \texttt{mons} array contains 12 pointers to string literals (\ie, 12 month names) in the program's data section. 
	\item[] \ding{183} The attacker knows the memory address of the global pointer \texttt{main\_server} at \texttt{0x80d6e14}, and reads the main server structure address pointed by \texttt{main\_server}, \ie, \texttt{0x871ae3c}.	
	The read operation is implemented by writing the address of the main server structure to the global writable location \texttt{G\_PTR}, and then transmitting the output via the exfiltration channel to the attacker side. 
	%cmd->server = main_server;  extern server_rec *main_server;  ==>  server_rec  ==> typedef struct server_struc {

	\item[] \ding{184} The attacker knows the offset of the field \texttt{ServerName} in the main server structure, which is \texttt{0x10} according to the binary of ProFTPd. Then, the attacker is able to calculate the address of  
	\texttt{main\_server->ServerName}, \ie, \texttt{0x871ae3c+0x10=0x871ae4c}.
	Given the memory address \texttt{0x80de0c8} of \texttt{ssl\_ctx}, \ie, the base pointer of a chain of 8 pointers to the private key, the attacker writes this address to \texttt{main\_server->ServerName} located at \texttt{0x871ae4c}. 
	\item[] \ding{185} Deference the base pointer \texttt{ssl\_ctx}, where the output is \texttt{0x874d7b8}. 
	The dereferencing operation dereferences the value currently located at \texttt{main\_server->ServerName}, by triggering the execution of the dereference gadget in line 4 of Listing~\ref{lst:sreplace}. The dereferenced value will be copied to a known position in the response buffer \texttt{resp\_buf}. Then, the attacker obtains the address \texttt{0x874d868} of \texttt{cert} by adding the offset \texttt{0xb0} to the dereferenced value \texttt{0x874d7b8} (D$_1$ in Fig.~\ref{fig:proftpd}). After that, the attacker copies the address of \texttt{cert} to \texttt{main\_server->ServerName} for the next iteration of deference. This step repeats 7 times (D$_1$$\sim$D$_7$ in Fig.~\ref{fig:proftpd}) following the dereference chain as shown in Fig.~\ref{fig:proftpd}. 
	The attacker needs to know the offset of the relevant field to the base address in each iteration, which can be derived from the binary or source code. Finally, the final address of the private key is obtained. 
	\item[] \ding{186} The attacker sequentially reads 8 bytes from the private key buffer via the information exfiltration channel constructed in the first step. This process repeats for 64 times to retrieve a total of 512 bytes data. 	
\end{enumerate}

%Gadget dispatchers are sequences of x86 instructions that equip attackers with the ability to repeat gadget invocations and, for each invocation, to selectively activate specific gadgets.

\subsection{Representative data-oriented attacks}\label{sec2:3}
%nginx stack buffer overflow Ubuntu 12.04 1 1
%CVE-2012-0809 sudo format string Ubuntu 12.04 1 0
%CVE-2009-4769 httpdx format string Windows XP SP3 4 1
%bugtraq ID: 41956 orzhttpd format string Ubuntu 9.10 1 1
%CVE-2002-1496 null httpd heap overflow Ubuntu 9.10 2 0
%CVE-2001-0820 ghttpd stack buffer overflow Ubuntu 12.04 1 0
%CVE-2001-0144 SSHD integer overflow Ubuntu 9.10 2 1
%CVE-2000-0573 wu-ftpd
In the seminal work of non-control data attacks~\cite{Chen:2005:NAR} and later FlowStitch~\cite{Hu:2015:DOA}, the authors have described more than 20 different data-oriented exploits (most of them are single-step DDM attacks). 
More recently, several research efforts have shown that data-oriented attacks pose serious threats to real-world programs.
%Overwriting security-critical data variables could cause escalation of privileges and leakage of sensitive data bytes~\cite{Hu:2015:DOA}. 
\begin{table*}[htb]
	\newcommand{\tabincell}[2]{\begin{tabular}{@{}#1@{}}#2\end{tabular}}
	\begin{center}
		\resizebox{16.0cm}{!}{
			
			\begin{tabular}{  >{\centering\arraybackslash}m{2.5cm} >{\centering\arraybackslash}m{1.5cm} >{\centering\arraybackslash}m{5.2cm}  >{\centering\arraybackslash}m{4.0cm} }
				\hline 
				\cellcolor{lightgray}{Targeted Application and Year} &  \cellcolor{lightgray}{Type}   & \cellcolor{lightgray}{Assumption/Requirement} & \cellcolor{lightgray}{Capability}  \\ %Weaknesses Strengths Privilege escalation
				\hline  \hline
				\emph{Chrome}~\cite{Jia:2016:CCS}, 2016  & DDM   & Identified security-critical variables, and arbitrary read/write capability & Bypass the same-origin policy\\
				\hline   
				\emph{Linux Page Table}~\cite{PTRand:NDSS:2017}, 2017  & DDM &  Kernel code writable,  and arbitrary read/write capability &  Bypass the kernel CFI \\
				\hline 
				\emph{InternetExplorer, Chrome}~\cite{Rogowski:EuroSP:2017}, 2017 &  DDM & Identified security-relevant variables, and arbitrary read/write capability  &  Information leakage, bypass the same origin policy, etc.\\
				\hline   
				\emph{Nginx}~\cite{Morton:EuroSP:2018}, 2018   & Multi-step DDM & Identified security-critical data structures, known unused portion of the data section, and arbitrary read/write capability  &Disable or degrade services, information leakage, etc.  \\
				\hline
				\emph{ProFTPd}~\cite{Hu16}, 2016   & DOP & Memory addresses of multiple involved data, identified gadgets/dispatchers, and arbitrary read/write capability  & Private key leakage w/ ASLR \\
				
				\hline
				\hline
			\end{tabular}
		}
	\end{center}	\vspace{-10pt}
	\caption{Recent data-oriented attacks pose serious threats against real-world programs.}
	\label{RepresentativeAttacks} \vspace{-10pt}
\end{table*}

Jia~\etal~\cite{Jia:2016:CCS} utilized data-oriented attacks to bypass the same-origin policy (SOP) enforcement in the Chrome browser. 
%The decision making variables related to SOP security policy checking in Chrome are identified by comparing differences in security-check functions across distinct execution traces recorded with the debugger. 
By manipulating the values of in-memory flags related to SOP security policy checking (which requires an arbitrary read/write privilege), the SOP enforcement can be undermined in Chrome. Davi~\etal~\cite{PTRand:NDSS:2017} showed that a data-only attack on page tables can undermine the kernel CFI protection. By manipulating the memory permissions in kernel page entries, the attack makes kernel code pages writable and subsequently enables malicious code injection to kernel space.

Rogowski~\etal~\cite{Rogowski:EuroSP:2017} introduced a new technique, called memory cartography, that an adversary can use to navigate itself at runtime to reach security-critical data in process memory, and then modify or exfiltrate the data at will.
They demonstrated the feasibility of data-oriented exploits against modern browsers such as Internet Explorer and Chrome, where possible attacks range from cookie leakage to bypassing the SOP. 
Morton~\etal~\cite{Morton:EuroSP:2018} demonstrated the potential threat of data-oriented attacks against asynchronous web servers (\eg, Nginx or Apache). By manipulating only a few bytes in memory, it is possible that an attacker re-configures a running asynchronous web server on the fly to degrade or disable services, steal sensitive information, and distribute arbitrary web content to clients. The attack consists of multiple steps (\ie, a multi-step DDM). It starts with locating the security-critical configuration data structures of the server and exposing their low-level state at runtime by leveraging memory disclosure vulnerabilities. Then, an adversary constructs faux copies of security-critical data structures into memory by exploiting memory corruption vulnerabilities. By redirecting data pointers to faux structures, a running web server instance can be re-configured by the attacker without corrupting the control-flow integrity or configuration files on disk. However, in the end-to-end exploits, authors in~\cite{Morton:EuroSP:2018} simulated the arbitrary write vulnerability in the recent version of Nginx, rather than exploiting a real-world vulnerability.   

Table~\ref{RepresentativeAttacks} summarizes these recent data-oriented attacks. 
Because existing CFI-based solutions are rendered defenseless under data-oriented exploits, such threats are particularly alarming. 
To construct a data-oriented exploit, attackers must have an in-depth knowledge of the vulnerable program's exact memory layout at runtime. 
In comparison to the DDM attack, a DOP attack requires non-trivial engineering efforts to chain gadgets for malicious effect.

\section{Defenses Against Data-Oriented Attacks} \label{defenseoverview}

In this section, we first  describe a three-stage model for data-oriented attacks, and a taxonomy of existing applicable defense techniques.
%Then, we give a brief overview of primary defenses in different stages. 
Then, we provide a comparative analysis of recent defensive approaches, particularly against data-oriented attacks.

\subsection{Three-stage model for launching data-oriented attacks}

\begin{figure*}[!h] %\vspace{-10pt}
	\centering
	\includegraphics[width=0.95\textwidth]{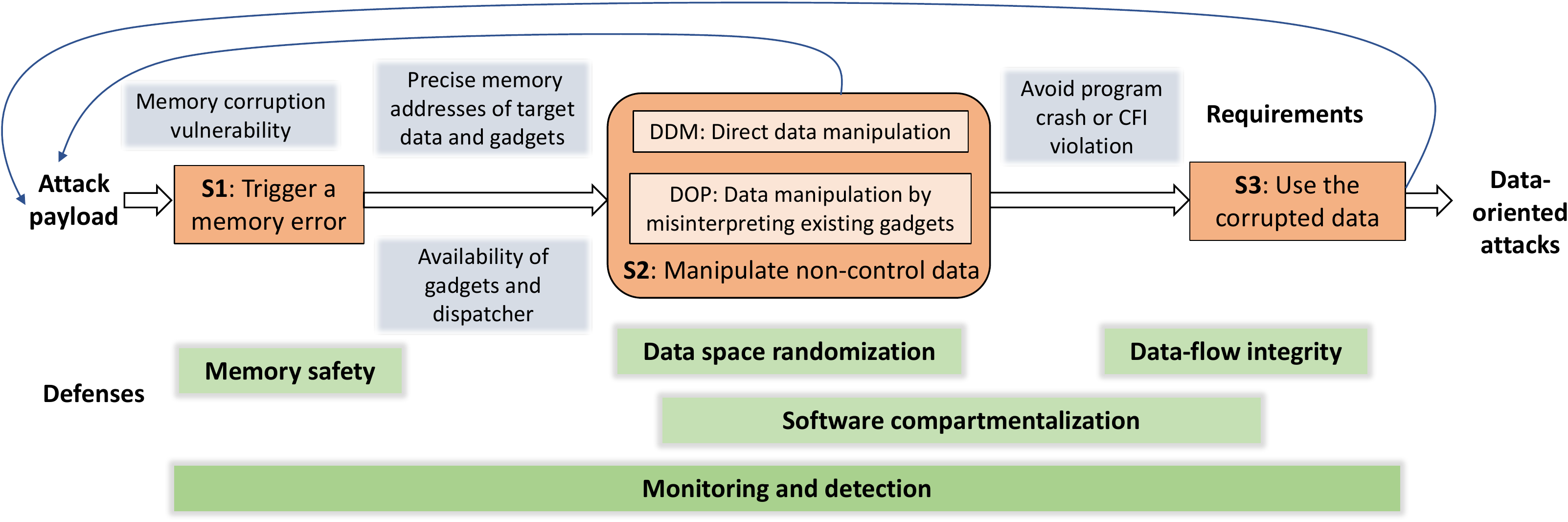} %\vspace{-10pt}
	\caption{Stages in data-oriented attacks and mitigation in different stages}  
	\label{fig:doamodel} \vspace{-10pt}
\end{figure*}

Fig.~\ref{fig:doamodel} illustrates the abstract view of three stages in data-oriented attacks. To launch such attacks, it starts with triggering a memory error of a vulnerable program (\ie, Stage S1), which empowers an attacker with control of the memory space, \eg, read/write capability. In Stage S2, the targeted non-control-data is modified (through either DDM or DOP). 
In Stage S3, the manipulated data variable is used and takes effect to change the default program behavior. Note that S3 does not necessarily happen immediately after the data manipulation. The back edges pointing from S3$\rightarrow$S1 and S2$\rightarrow$S1 indicate that an attacker may need to corrupt non-control-data multiple times to achieve the malicious goal. 

We discuss requirements in different stages (\ie, the threat model) that are essential to launching a successful DOP attack. The first three requirements apply for DDM exploits. 
 
\begin{enumerate}[*]
\item The {\em presence of a memory corruption vulnerability} (such as a buffer or heap overflow) in the target program, which allows attackers to modify the content of the application's memory (\ie, write capability). This is a reasonable assumption since low-level memory-unsafe languages (\eg, C/C++) are still in widespread use today due to interoperability and speed considerations, even though memory corruption vulnerabilities are an inevitable security weakness in these languages. 

\item Knowing the exact {\em location of target non-control data in memory}. Due to the wide deployment of exploit mitigation technologies such as DEP and ASLR, it is likely attackers need to first leverage memory disclosure vulnerabilities to circumvent the address space randomization~\cite{Morton:EuroSP:2018}. In this case, an exfiltration channel to achieve information leakage is needed (\ie, read capability), such as reading data from arbitrary addresses of the target program.

\item Knowing exactly the {\em transformation of an attack payload to the impact on memory space} of the target program. For example, a continuous buffer overflow may generate side effects that cause the program to crash. When launching a data-oriented exploit, attackers need to avoid any CFI violation and program crash.

\item {\em Availability of DOP gadgets} that are reachable by the memory corruption vulnerability, and triggerable by the attack payload.

\item {\em Stitchability of disjoint DOP gadgets}. A gadget dispatcher is needed to dispatch and execute the functional DOP gadgets. However, it is non-trivial to find gadget dispatchers in a program since they require loops with suitable gadgets and selectors controlled by a memory error.
\end{enumerate}

\subsection{Taxonomy of applicable defense techniques}

We briefly discuss defenses focusing on preventing these requirements from being satisfied at different points/stages. More generic memory corruption prevention mechanisms (in Stages S1 and S2) can be found in~\cite{Szekeres:2013:SEW, Larsen14, SoK:2018}.

\subsubsection{S1 Defense -- Preventing exploitation of memory errors}

\emph{Memory safety} enforcement is the first line of defense, which aims to prevent both spatial and temporal memory errors, such as buffer overflows and use-after-free errors. 
Memory-safe programming languages achieve this with built-in runtime bounds checks and garbage collection that make them immune to memory errors. In contrast, \emph{memory-unsafe} languages such as C/C++ lack built-in memory safety guarantees. Programs written in memory-unsafe languages therefore commonly exhibit memory errors that may make them vulnerable to runtime exploitation.
%They allow direct access to memory using pointers, which is a common cause of memory corruption. 
Enforcing all memory accesses staying within the bounds of intended objects would completely eliminate the pre-conditions for attacks that rely on gaining access to a prohibited area of memory. 
%Enforcing strict memory safety could fundamentally solve the memory corruption issue, which prevents all unexpected memory reads and writes. 
Despite considerable prior research in retrofitting memory-unsafe programs with memory safety guarantees, memory-safety problems persist due to an trade-off between effectiveness and efficiency: low-overhead approaches usually offer inadequate protection/coverage, while comprehensive solutions either incur a high performance overhead or provide limited backward compatibility~\cite{Szekeres:2013:SEW, Sinha:2018:ISCA}. 

\emph{SoftBound}~\cite{Nagarakatte09} and \emph{HardBound}~\cite{Devietti08} perform pointer bounds checks against metadata stored in a shadow memory area. 
SoftBound incurs an average performance overhead of 67\% in standard benchmarks. HardBound is a hardware-assisted scheme where the processor checks associated pointer bounds implicitly when a pointer is dereferenced. As the check is performed by hardware logic, the average performance overhead is reduced to around 10\%. Both schemes have a worst-case memory overhead of 200\%. \emph{Fat-pointer} schemes store the associated bounds metadata~\cite{Kuvaiskii17} together with pointers, \eg, by increasing their length~\cite{Necula02} or by borrowing unused bits from pointers~\cite{Kuvaiskii17}. But changing the representation of pointers in memory breaks both binary and source code compatibility. \emph{Code-Pointer Integrity} (CPI)~\cite{Kuznetsov:2014:CI} provides control-flow hijacking protection with a very low performance overhead (\eg, 8.4\% slowdown for C/C++ program). However, it only focuses on code-pointer checking without providing the complete memory safety.

\subsubsection{S2 Defense -- Providing a barrier to access to data or guess memory layout}
The purpose of S2 defenses is to mitigate the consequences of attacks in the presence of memory vulnerabilities. S2 defenses include \emph{software compartmentalization}~\cite{Wahbe93a, Erlingsson06, Mao11} and \emph{address space or data layout randomization}~\cite{Shacham:2004:EAR, Bhatkar08} techniques. They serve as the second line of defense, which creates a barrier for attackers trying to access target data or guess the memory layout.
 
Software compartmentalization isolates software components into distinct protection domains in order to limit the utility of existing memory errors (\ie, when the memory error and data to be manipulated exist in
different protection domains), but also limit the abilities of a compromised software component. For example, \emph{Software Fault Isolation} (SFI)~\cite{Wahbe93a} compartmentalizes software in a single address space by sandboxing untrusted modules into separate fault domains. This compartmentalization ensures that code in the fault domain is unable to directly access memory or jump to code outside the reserved portion of address space, but must interact with code outside it's domain through well-defined call interfaces.

\emph{Randomization} aims to hide attack targets by randomizing the layout of memory space~\cite{Larsen14, Giuffrida:Security:2012, Bigelow:2015:TRM}, layout of the code~\cite{Crane15}, layout of data~\cite{Bhatkar08} or the data itself~\cite{Cadar08} so that unauthorized access would lead to unpredictable behavior. 
In particular, data space randomization~\cite{Bhatkar08, Belleville:2018} aims to randomize the representation of data stored in program memory at runtime to make it unpredictable for unauthorized accesses, and thus reducing the possibility that attackers can leak security-critical memory addresses or manipulate the content of targeted data.  
ASLR~\cite{Shacham:2004:EAR} randomly chooses the base addresses of the stack, heap, code segment, and shared libraries. Data Space Randomization (DSR)~\cite{Bhatkar08} encrypts data stored in memory, rather than randomizing the location. 
\textcolor{notecolor}{Though strong randomization can stop memory corruption attacks with a high probability, the protection is confined to all data/addresses that are randomized/encrypted. In practice, to avoid a significant performance degradation, not all data/addresses are protected by randomization defenses~\cite{Szekeres:2013:SEW}.}
On the other hand, information leaks can undermine randomization techniques. In addition, data/address encryption based solutions are not binary compatible (\ie, protected binaries are incompatible with unmodified libraries)~\cite{Szekeres:2013:SEW}.
\begin{table*}[htb]
	\newcommand{\tabincell}[2]{\begin{tabular}{@{}#1@{}}#2\end{tabular}}
	\begin{center}
		\resizebox{18.0cm}{!}{
			\begin{tabular}{  >{\arraybackslash}m{2.7cm} >{\centering\arraybackslash}m{3.0cm} >{\centering\arraybackslash}m{2.5cm} >{\centering\arraybackslash}m{4.0cm} >{\centering\arraybackslash}m{2.5cm} >{\centering\arraybackslash}m{1.5cm} }
				\hline
				\cellcolor{lightgray}{Defense and Year} & \cellcolor{lightgray}{Stage} & \cellcolor{lightgray}{Approach} & \cellcolor{lightgray}{Security Guarantee}  & \cellcolor{lightgray}{Overhead}  & \cellcolor{lightgray}{General  Approach} \\ %Weaknesses Strengths
				\hline
				\hline
				\emph{YARRA}~\cite{Schlesinger11}, 2011 & S1 (Pointer safety)  &  Program instrumentation  & User-specified critical data & 400\%$\sim$600\% (whole program) &  \cmark \\
				\hline
				\emph{HardScope}~\cite{DBLP:NymanDZLPAS17}, 2018 &  S2 \& S3 (Compartmentalization) & Hardware extension  &  Context-specific memory isolation  & $\sim$3.2\% & \xmark\\
				\hline
				\emph{PrivWatcher}~\cite{Chen:2017:PrivWatcher}, 2017  &  S2 (Compartmentalization) &  Kernel modification  &  Protect process credentials data in Linux kernel   & $\sim$3\% (94\% in extreme cases) & \xmark\\
				\hline
				\emph{HDFI}~\cite{Song16}, 2016  & S2 (Compartmentalization)  &  Hardware extension  & Coarse-grained data-flow isolation  & $\sim$2\% & \xmark  \\
				\hline   
				\emph{PT-Rand}~\cite{PTRand:NDSS:2017}, 2017  & S2 (Randomization)  &  Kernel modification & Protect
				kernel page tables  & 0.22\% & \xmark \\
				\hline 
				\emph{DFI}~\cite{Castro06}, 2006  & S3 &  Program instrumentation  & Data-flow integrity  & $\sim$100\% &  \cmark\\
				\hline   
				\emph{CVI}~\cite{OEI:Sun:2018}, 2018   & S3 & Program instrumentation  &  Selective data-flow integrity  &   $\sim$2.7\%  & \xmark \\
				%	\hline
				%	\emph{PT-based Detection} & S1 -- S3  & Hardware-assisted morning   &  Branch behavior integrity   & $\sim$10\% tracing overhead \\ %and online control-flow reconstruction 
				\hline
				\hline
			\end{tabular}
		}
	\end{center}	\vspace{-10pt}
	\caption{Comparison of defensive mechanisms against data-oriented attacks}
	\label{relatedwork:comparison} \vspace{-15pt}
\end{table*}

\subsubsection{S3 Defense -- Preventing/detecting use of corrupted data}

Data-Flow Integrity (DFI)~\cite{Castro06} mitigates data corruption before the manipulation takes effect. Before each read instruction, DFI ensures that a variable can only be written by a legitimate write instruction which can be derived by reaching definitions analysis (\ie, for each value read instruction, it statically computes the set of write instructions that may write the value). However, DFI usually overestimates the set of valid write instructions since the set is statically determined without runtime information. Moreover, Software-based DFI incurs a high performance overhead~\cite{Hu16} due to the frequent read instruction checking. Intra-procedural DFI incurs 44\% and inter-procedural DFI incurs 103\% runtime performance overhead, respectively, and approximately 50\% space overhead for instrumentation~\cite{Castro06}.
Hardware-based DFI, \eg, HDFI~\cite{Song16}, is efficient, but limited by the number of simultaneous protection domains it can support. 

Depending on the granularity of compartmentalization and the boundaries of the security domain, software compartmentalization can also function as a defense in S3. It can prevent the use of corrupted data. For example, when a corrupted pointer is referencing memory in another protection domain, it thwarts the dereference operation.

%stop attacks in their later phases.
Szekeres~\etal~\cite{Szekeres:2013:SEW} provide a systematic overview of memory corruption attacks and mitigations. They highlighted that though a vast number of solutions have been proposed, memory corruption attacks continue to pose a serious security threat. Real-world software exploits are still possible because currently deployed defenses can be bypassed. Program anomaly detection complements the aforementioned mitigation techniques, and serves as the last line of defense against data-oriented attacks.
%low-level in-lined defences.
%existing solutions: inlined reference monitors
%Different from~\cite{Szekeres:2013:SEW}, 
As shown in Fig.~\ref{fig:doamodel}, passive monitoring based program anomaly detection has the potential to detect anomalous program behaviors exhibited in all the three stages of data-oriented attacks. We will discuss its detection capabilities in details in Sections~\ref{feasibility-overview} and~\ref{sec5:detection}.

\subsection{Defense mechanisms against data-oriented attacks}\label{defensivemechanisms}

%\subsection{S1 Defenses}
In addition to generic memory corruption prevention mechanisms, a number of detection and prevention techniques specially focusing on data-oriented attacks have been proposed in the literature. In this section, we discuss these defenses. 
%XXX aims to prevent XXX, XXX detects it.

\emph{YARRA}~\cite{Schlesinger11} is a C language extension that validates a pointer's type for \emph{critical data types} annotated by developers, which is an S1 defense. It guarantees that critical data types are only written through pointers with the given static type. YARRA is suitable for hardening access to isolated pieces of critical data, such as cryptographic keys stored in program memory at runtime. However, when applied for the whole program protection, it incurs a performance overhead in the order of 400\%$\sim$600\%. In addition, YARRA \textcolor{notecolor}{relies on the programmers' manual annotations}, which is undesirable for complicated programs.

\emph{HardScope}~\cite{DBLP:NymanDZLPAS17} is a hardware-assisted variable scope enforcement approach to mitigate data-oriented attacks by introducing intra-program memory isolation based on C language variable visibility rules derived during program compilation. On each memory access (\ie, load/store), HardScope enforces that the memory address requested is in the accessible memory areas. 
Nyman~\etal~\cite{DBLP:NymanDZLPAS17} demonstrated the effectiveness of HardScope for the RISC-V open instruction set architecture, by introducing a set of seven new instructions. HardScope instructions are instrumented at compile-time, and memory access constraints are enforced at runtime. It shows that HardScope has a real-world performance overhead of 3.2\% in embedded benchmarks.
Although HardScope significantly reduces the usefulness of DOP gadgets and thwarts Hu~\etal~\cite{Hu16}'s example attacks, HardScope cannot guarantee the absence of DOP gadgets in arbitrary programs.

%The authors highlighted that though HardScope significantly reduces the number of available DOP gadgets, it is still not possible to defeat all DOP attacks in arbitrary programs. 

\emph{PrivWatcher}~\cite{Chen:2017:PrivWatcher} is a framework for monitoring and protecting the integrity of process credentials (\ie, \texttt{task\_struct} that describes the privileges of a process in the Linux kernel) against non-control data attacks. It involves a set of kernel modifications including relocating process credentials into a safe region, code instrumentation and runtime data integrity verification, in order to provide non-bypassable integrity assurances. It ensures the Time of Check To Time of Use (TOCTTOU) consistency between verification and usage contexts for process credentials by adopting a dual reference monitor model. The authors implemented the PrivWatcher prototype in Ubuntu Linux. The experiment results show that PrivWatcher incurs an overhead less than 3\%. But it incurs more than 94\% overhead for applications that involve installing new \texttt{task\_struct} structures to processes.

%prevents exploitable kernel-level code from modifying process credentials data in Linux 
%is a framework for monitoring and protecting the integrity of process credentials and their usage contexts from memory corruption attacks.

\emph{Hardware-Assisted Data-flow Isolation} (HDFI)~\cite{Song16} extends the RISC-V architecture to provide an instruction-level isolation by tagging each machine word in memory (also known as the tag-based memory protection). The one-bit tag of a memory unit in HDFI is defined by the last instruction that writes to this memory location. At each memory read instruction, HDFI checks if the tag matches the expected value. However, unlike software-enforced DFI, HDFI only supports two simultaneous protection domains.

%Tagged memory architectures, \eg, \emph{lowRISC}~\cite{lowRISC}, in general can be used to implement sophisticated compartmentalization strategies. \emph{Intel Memory Protection Keys} (MPK)~\cite{MPK} provides a hardware support for page granularity tagged memory. MPK allows a process to assign memory pages to one of 16 distinct protection domains and selectively enable or disable access to each region independently.

%tag-based as opposed to the content-based?
% they use metadata tags, stored in a region separate from program data, to compare and verify access validity. This, in turn, requires (explicit or implicit) out-of-band fetching and processing of metadata

Davi~\etal~\cite{PTRand:NDSS:2017} presented a data-oriented attack against kernel page tables to bypass CFI-based kernel hardening techniques, and subsequently attackers can execute arbitrary code with kernel privileges. To mitigate the threat, they proposed \emph{PT-Rand}, which randomizes the location of page tables to prevent attackers from manipulating page tables by means of data-oriented attacks. Evaluation results show that PT-Rand on Debian only incurs a low overhead of 0.22\% for common benchmarks.
However, it is still possible attackers undermine these schemes if the secret information (\eg, randomization secret) is leaked or inferred~\cite{DBLP:NymanDZLPAS17}. 

%\subsection{S3: Defense}

\emph{CVI} (Critical Variable Integrity)~\cite{OEI:Sun:2018} verifies define-use consistency of critical variables for embedded devices. The define-use consistency is defined as the property that the value of a variable cannot change between two adjacent define- and use-sites. 
After identifying critical variables (either automatically identified or manually annotated), the compiler inserts instrumentation at all the define- and use-sites for these critical variables, to collect values at runtime and send them to an external measurement engine. CVI checking compares the current value of a variable at every
use-site, and the recorded value at the last legitimate define-site. However, like DFI~\cite{Castro06}, CVI is based on compile-time instrumentation and frequent runtime checking, which incurs a high overhead for the complete protection. 
 
%On the other hand, correlated branches exhibit runtime control-flow patterns during program execution. ~\cite{DBLP:NymanDZLPAS17}

Table~\ref{relatedwork:comparison} compares representative data-oriented attack specific defensive mechanisms. 
%Most approaches are recently proposed except DFI~\cite{Castro06}. 
PrivWatcher~\cite{Chen:2017:PrivWatcher}, HDFI~\cite{Song16}, PT-Rand~\cite{PTRand:NDSS:2017}, and CVI~\cite{OEI:Sun:2018} protect specific non-control data.
HardScope~\cite{DBLP:NymanDZLPAS17} can protect against all DOP attacks that violate variable visibility rules at runtime. However, it requires developer assistance in certain settings. The main drawback of HarScope and other solutions based on new hardware extensions~\cite{Song16} is the high bar for deployment.
They cannot be directly applied to protect user-space applications against general data-oriented attacks, in particular DOP attacks. On the hand, the two general approaches DFI~\cite{Castro06} and YARRA~\cite{Schlesinger11} incur a high performance overhead at runtime.

\section{Overview of Detectability Based on Control-flow Tracing}\label{feasibility-overview}

This and the next sections describe our effort of understanding the data-oriented attacks from the defense perspective. In particular, we consider the scenario where one can efficiently trace control flows. With new hardware development, namely the Intel Processor Trace (PT)\footnote{PT is a low-overhead hardware feature on Intel CPUs that enables the construction of the complete control flows during program execution.}, runtime control-flow tracing becomes realistic. Real-time control-flow monitoring without significantly slowing down the program execution can be widely deployed. 
\textit{Control-flow based detectability analysis may sound like an oxymoron, because data-oriented attacks do not violate control-flow integrity. However, this superficial view is incomplete}. There exist cases where the anomalies may be detected by correlating or aggregating multiple (legal) control-flow observations.

%---clnote---later check where to put this
%In 2006, Zhuang~\etal~\cite{Zhuang:MICRO:2006} showed that many non-control data attacks cause changes in control-flows. In their experiments, on average 49.4\% of memory tamperings change program control-flows. 

%---clnote---later check where to put this
The \textit{necessary condition} for detecting a data-oriented attack using control-flow tracing is that the attack directly or indirectly affects the flow of a program's execution. Typically, uses of non-control data in a program can be classified either as predicate uses or non-predicate uses, such as computation uses. A predicate-use directly affects the control flow. While a non-predicate use may affect the computation or the output of a program~\cite{Rapps:1985:TSE}. 
Suppose that PT-style control-flow tracing (namely, recording indirect control transfers and conditional branches) is enabled. In what follows, we categorize the cases where data-oriented attacks are impossible to detect (\ie, undetectable cases) and cases where such attacks may be detected (\ie, detectable cases) by control-flow tracing. For the detectable cases, we provide more specific experimental observations in Section~\ref{sec5:detection}. We report our overhead and performance measurement in Section~\ref{sec5:overhead}.

%\noindent
\subsection{Undetectable cases} 
Detecting data-oriented attacks using control-flow tracing requires that an attack manifests incompatible/unusual control-flow behaviors, \eg, incompatible branch behaviors or frequency anomalies. However, when a manipulated variable is only used for computation or output (\ie, non-predicate use), and the exploit does not incur any side effect on control transfers, such an attack is undetectable by PT control-flow tracing. We list typical undetectable cases, which are mainly direct data manipulation (DDM) attacks.
 
\begin{enumerate}[*]

	\item Corrupting user identity for privilege escalation: Simply corrupting user identify data (\eg, \texttt{UID}) may lead to a compromise of the root privilege. However, for an undetectable data manipulation with the privilege escalation, an attacker usually goes after malicious actions once obtaining the new privilege, \eg, launching a shell. Such malicious actions can be easily detected by control-flow tracing.
	\item Corrupting configuration data: Corrupting configuration data via format string vulnerabilities may evade PT-based detection, since format string vulnerabilities allow a single memory write without a side effect on control-flow behavior.
	\item Constructing exfiltration channels for information leakage: Attackers exploit an existing information outlet (also known as sink functions such as \texttt{printf} or \texttt{send}) for information leakage by replacing the pointer value of the outlet function's parameter with the address of the data to be exfiltrated. Such an attack may not incur any anomalous control-flow behavior. 
\end{enumerate}

%In this section, we discuss the possibility of detecting data-oriented attacks through program anomaly detection based on the fast Intel Processor Trace (PT) technology, and study the limitations of such protection.

\subsection{Possibly detectable cases} 

We observe that data-oriented attacks can potentially cause three types of anomalous control-flow behaviors. 

\begin{enumerate}[*]  

	\item \emph{Incompatible branch behavior}: Manipulating a predicate-use variable (\eg, decision-making data) can change the default branch behavior of a program.
	If there exist two correlated conditional branches that are data-dependent on the manipulated variable before and after the data manipulation site, it is likely the data manipulation incurs incompatible branch behaviors that can be detected by control-flow tracing. For example, in Listing~\ref{lst:authen}, the conditional branches in lines 5 and 16 are correlated, since they both are data-dependent on the variable \texttt{authenticated}. 
	Suppose \texttt{authenticated} is corrupted at line 6 and there is no write	to \texttt{authenticated} after the data manipulation. 
	As a result, \texttt{while(!authenticated)} in line 5 returns \texttt{true}, but \texttt{if(authenticated)} in line 16 also returns \texttt{true}.
	We observe an incompatible branch behavior, which is detected when the corrupted variable \texttt{authenticated} is used in line 16 (\ie, in Stage 3 of the attack).

	In addition, a continuous buffer overflow may generate side effects on control-flow behavior, which could result in an incompatible control-flow path observable in Stage S2. For example, though the target buffer is not used for predicate-use, some decision-making variables close to the buffer may be inevitably corrupted. We manually analyzed 14 vulnerable programs in a test suite for buffer overflows~\cite{SRD}, and found that 5 out of 14 overflows cause side impacts on decision-making variables (\ie, involved in predicate expressions). 
	
	In Section~\ref{sec5:branchbehavior}, we experimentally characterize the branch correlation behaviors of ProFTPd under the DOP attack and report our findings.

	\item \emph{Macro-level interaction frequency anomaly}: In DOP (also BOP) attacks, an attacker normally needs to interact with a vulnerable program to repeatedly corrupt variables to achieve the attack purpose and avoid segmentation faults. This attack activity inevitably results in frequency anomalies during the client-server interaction, which can also be captured by control-flow tracing. For example, in the ProFTPd DOP attack introduced in Section~\ref{sec2}, an attacker needs to send a large number of FTP commands with malicious inputs to the ProFTPd server to corrupt the program memory repeatedly. 
	%In the BOP's spawning a shell attack~\cite{Ispoglou:2018:BOP}, it requires the attacker to send 5 malicious payloads to the vulnerable Nginx server. 
	
	\item \emph{Micro-level control-flow frequency anomaly}: Short control-flow paths may exhibit unusual execution frequencies. For instance, corrupting variables which directly or indirectly control loop iterations can cause such frequency anomalies.
	Micro-level control-flow frequency anomalies may be observed in different stages of data-oriented attacks. In addition, control-flow bending (CFB) attacks~\cite{Carlini:2015:CBE} and resource wastage attacks~\cite{Baliga:2007:SP} may also lead to unusual control-flow frequencies.

	In Section~\ref{sec:macro-frequency}, we experimentally compare the frequency differences of both the macro-level interactions and micro-level control-flows in normal ProFTPd executions and under a DOP attack.
	
	%In Section~\ref{sec:micro-frequency}, we experimentally compare the frequency differences of the micro-level control-flows in normal ProFTPd executions and under a DOP attack.
	
\end{enumerate}

\section{Characterization of DOP Behaviors with PT}\label{sec5:detection}

In this section, we provide more concrete observations about DOP's detectability based on control-flow tracing. 
Specifically, we experimentally measure the execution patterns of the ProFTPd DOP attack on a computer with Processor Trace (PT) enabled. The ProFTPd DOP attack is the only publicly demonstrated end-to-end DOP attack~\cite{Hu16}. Our goal is to illustrate the DOP's non-negligible impact on PT traces, as opposed to proposing a specific anomaly detection system. 
% 
%We first empirically investigate the feasibility of using branch correlation for detection of incompatible branch behaviors. Then, we conduct case studies to assess the PT-based detection capability against data-oriented attacks. 
%

We organize our experiments according to the detectable cases discussed in Section~\ref{feasibility-overview}. Our experiments mainly aim to answer the following research questions (RQs).
\begin{enumerate}[]
	\item {\bf RQ1:} Does the ProFTPd DOP attack exhibit any incompatible branch behavior?	How feasible of using branch correlation for detecting
incompatible branch behavior in real-world programs. (Section~\ref{sec5:branchbehavior})  %(Section~\ref{sec:branchbenchmarks})?  
	%	How often do correlated branches arise in real-world programs?
%	\item {\bf RQ2:} 
	
	\item {\bf RQ2:} Does the ProFTPd DOP attack exhibit any frequency anomalies that can be captured by PT tracing? (Section~\ref{sec5:dopdetection})  
	 
	\item {\bf RQ3:} What is the runtime performance overhead of PT tracing when monitoring the ProFTPd server? (Section~\ref{sec5:overhead})
\end{enumerate}

%Intel(R) Core(TM) i7-6700K CPU Linux kernel 4.3
We ported the original ProFTPd DOP attack to a 32bit Ubuntu 16.04 with PT support. The attack heavily relies on the precise knowledge of the library and memory layout. In particular, we modified the original metasploit module and created an automated script to scan the ProFTPd binary, which automatically locates the targeted memory addresses. This porting effort was non-trivial\footnote{The automated script of launching the ProFTPd DOP attack, our LLVM-based analysis tools, and all traces of the case studies are provided at https://goo.gl/5hmaZH}. 
%{\bf Long, if Hans' code is on github anonymously, then we should included it here.}
%\footnote{\scriptsize We share the automated script at XXXXX}
Our testing platform for the experiments is a desktop computer with an Intel i7-8700 CPU and 32GB of RAM unless otherwise specified. We collected control-flow traces using the Linux \texttt{Perf-PT} tool.

%We discuss practical deployment issues for hardware-assisted tracing in Section~\ref{sec7:discussion}.
%
% To characterize branch behaviors and micro-level control-flow frequencies, we configured \texttt{perf} to only trace the vulnerable \texttt{sreplace} function.

\subsection{Tracing indirect and conditional branches with PT}

%{\bf Long, let's show examples of the indirect branch and conditional branch, which would help readers understand the rest of this section. otherwise, it'd be tough for them}

In this section, we briefly explain what control-flow information PT records, which is needed to understand our experimental results.
PT tracing is at the instruction level, which is much more fine-grained than system- or function-call level tracing. To capture control-flow information, PT records target addresses of indirect branches  (\ie, TIP packets/events for the indirect \texttt{call}, indirect \texttt{jmp}, and \texttt{ret}) and taken/non-taken decisions of conditional direct branches (\ie, TNT packets/events). 
The trace format in PT is highly compressed to achieve efficient logging at runtime. For example, it uses one bit to indicate taken or not-taken for a conditional branch. %PT was originally designed for detailed offline debugging and failure diagnosis. 

Recent research has shown PT's applicability for on-line security enforcement to defend against control-flow attacks (referred to as dynamic CFI solutions).
{\sc Griffin}~\cite{Ge:2017:GGC} is an operating system mechanism (running in the kernel) that leverages the PT feature to enforce CFI policies. 
%It focuses on the effective on-line enforcement and assumes that CFI policies are given. 
{\sc PT-CFI} \cite{Gu:2017:PTB} and {\sc FlowGuard}~\cite{Liu:HPCA:FlowGuard} are two backward-edge control-flow violation detection systems using PT tracing.
To address the over-approximation problem of control targets in forward-edge CFI, {\sc PittyPat}~\cite{Ding:Security:2017} utilizes PT to track basic block execution to compute the legal control transfer target sets through runtime path-sensitive point-to analysis. But {\sc PittyPat} still makes approximations, as an incomplete execution context is used in its points-to analysis. 
$\mu$CFI~\cite{Hu:2018:EUC} improves {\sc PittyPat} by recording full execution context with PT to perform an accurate points-to analysis, and thus getting a unique code target for each indirect control-flow transfer. 

%Though it significantly reduces the amount of allowed targets, 

However, none of these existing work investigates the PT-based detection against data-oriented attacks. In what follows, we look into the possibility of PT-based detection against DOP attacks, which can also be applied to detect DDM attacks.
%The observed incompatible branch behavior and micro-level control-flow frequency anomaly
%Specifically, we present specific cases where PT tracing helps detect data-oriented attacks. 

\subsection{RQ1: Branch behavior patterns} \label{sec5:branchbehavior}

%\subsubsection{\textbf{Branch Behaviors in ProFTPd DOP}}\label{sec:branch-exp-1}

%{\bf Long, what's coarse grained? remove?}

We first utilize correlations among
branches to spot incompatible branch behavior in the ProFTPd DOP attack. Then, we characterize the branch correlation in benchmark programs to show the feasibility and generality of this approach.  

%characterize the branch correlation behaviors of ProFTPd under the DOP attack and report our findings
% present the detection of incompatible branch behavior  

We define the {\em subsume} and {\em mutually exclusive} deterministic relations among branch conditions/predicates. If a branch condition BR$_i$ returns \texttt{true}/\texttt{false}, another branch condition BR$_j$ must return \texttt{true}/\texttt{false}, we say BR$_i$ subsumes BR$_j$.
%(which is a one-way relationship, \ie, BR$_j$ does not subsume BR$_i$ unless they have same conditions) 
BR$_i$ and BR$_j$ are mutually exclusive, if they always return different results. For example, (x>10) subsumes (x>5), and (x>10) and (x<5) are mutually exclusive.

%In this specific ProFTPd attack, 
\subsubsection{\textbf{Identifying branch correlation in ProFTPd DOP}} 
\begin{figure}[h] %\vspace{-10pt}
	\centering
	\includegraphics[width=0.99\columnwidth]{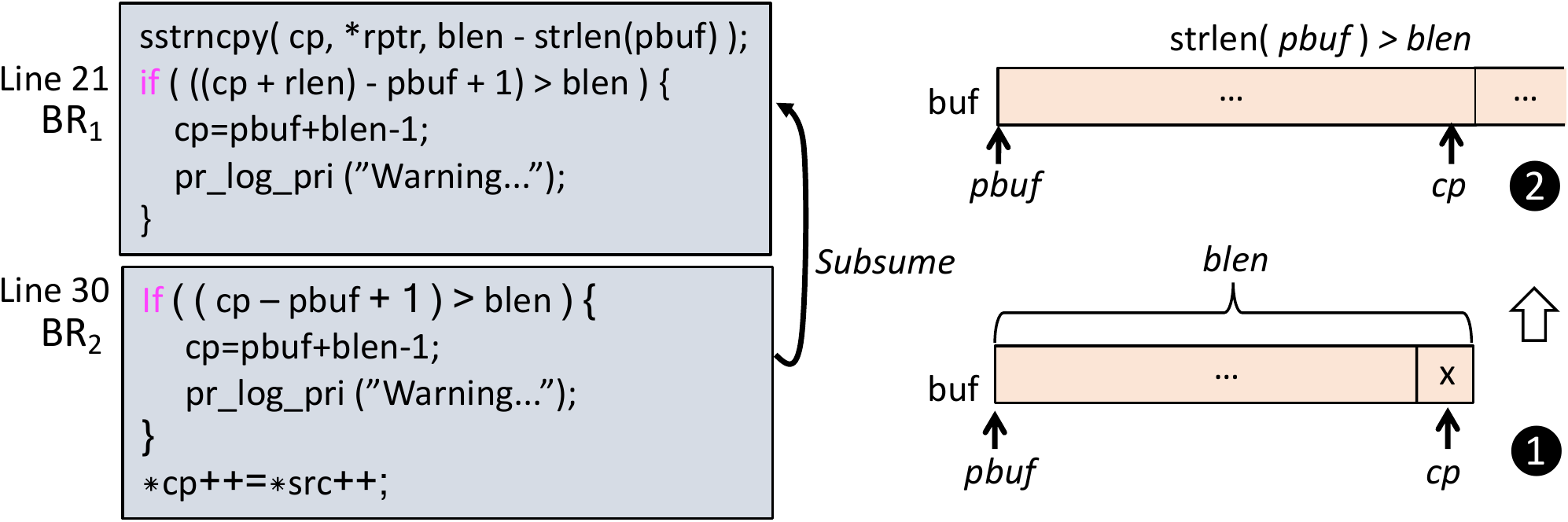} %\vspace{-10pt}
	\caption{Branch correlation and an incompatible branch behavior in the ProFTPd DOP attack}  
	\label{fig:corruption} \vspace{-8pt}
\end{figure}

For ProFTPd's vulnerable function \texttt{sreplace} (Listing~\ref{lst:sreplace}), the conditional branches in lines 21 and 30 of Listing~\ref{lst:sreplace} are correlated. Both are data dependent on the same variables \texttt{cp}, \texttt{pbuf}, and \texttt{blen}. Fig.~\ref{fig:corruption} shows these two correlated branches, denoted as BR$_1$ and BR$_2$, respectively. 

Take the example in Fig.~\ref{fig:corruption}, we could use the satisfiability modulo theories (SMT) solver Z3~\cite{Z3:2008} to determine any subsume or mutually exclusive relationship between correlated branches. If the SMT solver could not find a solution for 
the combined constraints (BR$_2$) and \texttt{Not}(BR$_1$) (\ie, the logical not of BR$_1$), we derive that BR$_2$ subsumes BR$_1$.
%If the predicate coverage of branch BR$_2$ is no larger than the predicate coverage of branch BR$_1$, we
In order to derive the subsume relationship from BR$_2$ to BR$_1$, we first add the predicate BR$_2$ as a constraint into Z3 solver. However, \texttt{cp} gets redefined in line 31 (Listing~\ref{lst:sreplace}), which is true-control-dependent on the branch in line 30. Thus, we need to replace BR$_2$ with the statement in line 31 as the constraint, \ie, \texttt{cp==pbuf+blen-1}. Then, we add the constraint \texttt{Not(((cp+rlen)-pbuf+1)>blen)} to the solver. Because the variable \texttt{rlen} is the length of a non-null string (derived from the source code), we add $rlen>0$ as an additional constraint. 
At the end, the solver returns \texttt{unsat} (satisfiable), and thus we derive that BR$_2$ subsumes BR$_1$, \ie, if BR$_2$ returns \texttt{true}, BR$_1$ should also take the \texttt{true} branch. 

\subsubsection{\textbf{Incompatible branch behavior in ProFTPd DOP}}
During the course of the ProFTPd DOP attack, the attacker first triggers the memory corruption error in \texttt{sreplace}, 
by filling up \texttt{buf} (where BR$_2$ returns \texttt{true}) and overwriting \texttt{buf}'s terminator with a non-zero byte (in line 33 of Listing~\ref{lst:sreplace}), as shown in \ding{182} in Fig.~\ref{fig:corruption}. Since \texttt{buf}'s last character is a non-zero value, it becomes a non-terminated string. As a result,  \texttt{strlen(pbuf)}>\texttt{blen} (\ding{183} in Fig.~\ref{fig:corruption}), which enables the attacker to corrupt the local variables such as \texttt{cp} and \texttt{blen} in line 20. To bypass the overflow checking in lines 21-27 in the following iterations, the attacker needs to make sure that the predicate in line 21 returns \texttt{false}. 
From the PT trace, we could observe the predicate in line 21 (\ie, BR$_1$) takes the \texttt{false} branch. 
Since BR$_2$ subsumes BR$_1$, and BR$_2$ has taken the \texttt{true} branch, the runtime branch behavior of BR$_1$ and BR$_2$ are incompatible.
Note that the observed incompatible branch behavior is not specific to DOP attacks. For any DDM attack against the ProFTPd, as long as the attacker exploits the same vulnerability (CVE-2006-5815~\cite{ProFTPD}), we could observe this incompatible branch behavior.

%can easily detect the anomaly using PT tracing. 

%{\bf Long, please remember discuss (2-3 sentences) later how the limited scalability of SMT impacts this approach. i mean how realistic is this?  that discussion is important for SoK - we don't need to advocate for any defense}

%{\bf Long, this transition between the 2 paragraphs -- above and below -- is unclear to me. can you add an opening sentence to explain what's this paragraph for?}

%{\bf Long, I'm not sure how the above paragraph is related to previous paragraphs. please help smooth this out}

\subsubsection{\textbf{Branch correlation in benchmarks}} \label{sec:branchbenchmarks}

%However, it is non-trivial to derive the relations between two branches with complex predicate expressions.
Our case study of the ProFTPd DOP attack demonstrates that correlated branches are useful for identifying incompatible branch behaviors.
To reflect the possibility of data-oriented attacks to manifest incompatible branch behaviors, we characterize the prevalence of branch correlation in benchmark programs.

%We first explain the types of branches we aim to measure.  There are two types of joint data dependent branches that we aim to record:

In order to capture branch correlations with arbitrary predicate expressions, we define the correlated branches to be branches that have joint data dependency, including {\em i)} \emph{direct} data-dependent branches that share at least one common predicate variable; and {\em ii)} \emph{indirect} data-dependent branches that use different predicate variables, but they are data-dependent on at least one common variable. We also define the {\em simple forms of branches}, where a branch predicate simply compares a variable with a constant value, \eg, the conditional branches in lines 5 and 16 in  Listing~\ref{lst:authen} are simple branches. Previously, Zhang~\etal~\cite{Zhuang:MICRO:2006}~used correlation among simple forms of branches to detect infeasible program paths caused by memory corruptions.  

%instruction is  simple, if it only has two backward data-dependent instructions (namely, \verb1icmp1 and \verb1load1). 
We developed a
branch correlation analysis tool based on LLVM~\cite{LLVM}. Our tool handles \emph{inter-procedural} branch correlations and \emph{arbitrary predicate expressions}.
We performed our branch correlation analysis on eight programs. They include four Linux utility programs (\texttt{flex}, \texttt{grep}, \texttt{gzip}, \texttt{sed}) from the Software-artifact Infrastructure Repository (SIR)~\cite{sir} and four vulnerable programs (\texttt{wu-ftpd}, \texttt{orzhttpd}, \texttt{ghttpd}, \texttt{sudo}) from the FlowStitch benchmarks~\cite{Hu:2015:DOA}. 
%{\bf Long, please explain coarse-grained, before you use that term. what do you mean? If we don't have fine-grained counterpart, then there's no need to differentiate the granularity.}

Table~\ref{tab:brannchanalysis} reports the branch correlation results. Overall, 24\% of the branches exhibit simple forms of conditional predicates, and 18\% of the branches with simple forms are correlated (including direct and indirect correlations). For the direct correlations with simple forms (\ie, correlated branches use the same predicate variable), we used the SMT logic solver Z3~\cite{Z3:2008} to determine any "subsume" or "mutually exclusive" relationship. Our results show that for a limited number of branch correlations (around 2\%), we can directly derive the deterministic correlation relationship (denoted as "directly derivable simple BRs" in Table~\ref{tab:brannchanalysis}). It also reflects that the branch correlation analysis in~\cite{Zhuang:MICRO:2006} has a very limited coverage. 

\begin{table}[!htb]
	%\vspace{-11pt}
	\begin{center}
		\resizebox{9.0cm}{!}{
			\begin{tabular}{  >{\centering\arraybackslash}m{1.2cm} >{\centering\arraybackslash}m{1.1cm} >{\centering\arraybackslash}m{1.5cm} >{\centering\arraybackslash}m{1.3cm} >{\centering\arraybackslash}m{1.5cm} >{\centering\arraybackslash}m{2.15cm} }
				\hline 	
				\cellcolor{lightgray}{Application} & \cellcolor{lightgray}{Total BRs}  & \cellcolor{lightgray}{Correlated BRs} & \cellcolor{lightgray}{Simple BRs} & \cellcolor{lightgray}{Correlated Simple BRs} & \cellcolor{lightgray}{Directly Derivable Simple BRs} \\
				\hline \hline
				%\hline
				\texttt{flex} & 1142 & 813 (71\%) & 557 (49\%) & 356 (31\%) & 62 (5\%)\\
				\hline
				\texttt{grep} & 1664  &1456 (87\%) & 278 (17\%) & 216 (13\%) & 32 (2\%)\\
				\hline
				\texttt{gzip} & 737  & 533 (72\%) & 241 (9\%) & 169 (23\%) & 60 (8\%)\\
				\hline
				\texttt{sed} & 1081 & 1017 (94\%) & 172 (16\%) & 142 (13\%) & 12 (1\%)\\
				\hline
				\texttt{wu-ftpd} & 2781 & 1943 (70\%) & 688 (25\%) & 398 (14\%) & 84 (3\%)\\
				\hline
				\texttt{orzhttpd} & 35 & 23 (66\%) & 13 (37\%) & 7 (20\%) & 0 (0\%)\\
				\hline
				\texttt{sudo} & 675 & 499 (74\%) & 163 (24\%) & 116 (17\%) & 0 (0\%)\\
				\hline
				\texttt{ghttpd} & 107 & 90 (84\%) & 16 (15\%) & 11 (10\%) & 0 (0\%)\\
				\hline
				Average & 1028 & \cellcolor{lightgray}{77\%} & \cellcolor{lightgray}{24\%} & \cellcolor{lightgray}{18\%} & \cellcolor{lightgray}{2\%}\\
				\hline 	\hline
				%\vspace{-16pt}
			\end{tabular}
		}
	\end{center}\vspace{-8pt}	
	\caption{Branch correlation analysis in benchmarks}
	\label{tab:brannchanalysis}
	\vspace{-8pt}	
\end{table}

We observed 77\% of the branches have at least one correlated branch, \ie, given BR$_i$, we can find at least one branch BR$_j \neq$ BR$_i$, where BR$_i$ and BR$_j$ have joint data dependency. 
Though our analysis only captures the coarse-grained
branch correlations as opposed to the deterministic relationships among branches, this result suggests the prevalence of branch correlations with complex predicate expressions in a program, which can be potentially used as checkpoints to detect incompatible branch behaviors (\eg, first using symbolic execution or dynamic analysis techniques to identify deterministic relationships among these correlated branches).

\subsection{RQ2: Frequency patterns} \label{sec5:dopdetection}
 
We characterize how DOP impacts the quantitative behaviors of ProFTPd by comparing the frequency distributions in DOP attacks and normal executions. We conducted two sets of comparisons, {\em i)} on macro-level interaction frequencies and {\em ii)} on micro-level control-flow frequencies. Both are defined in Section~\ref{feasibility-overview}.

%[no behavior instance for macro-level interactionsm, only for micro-level control-flow frequencies] When analyzing the macro-level interactions between a FTP server and a client, we define each connection to be a behavior instance, following the approach in~\cite{Shu:2015:USP}. 
%Thus, we adopt simple feature extraction ($n$-gram) and dimension reduction (PCA) approaches. We describe our measurements and findings next.

\subsubsection{\textbf{Macro-level interaction frequencies}}
\label{sec:macro-frequency}

To detect interaction frequency anomalies under
the DOP attack, we derived the FTP commands sent from clients by tracing control-flow transfers
of the FTP command dispatcher function \texttt{\_dispatch} in the ProFTPd server program. PT captures the control-flow transfers from \texttt{\_dispatch} to different command handlers, \eg, \texttt{core\_cwd} indicates that the command \texttt{CWD} (\ie, change working directory) has been received. 

To characterize the baseline FTP interaction frequencies, we used the LBNL-FTP-PKT~\cite{LBNL-FTP-PKT} dataset. It contains all incoming anonymous FTP connections to public FTP servers at the Lawrence Berkeley National Laboratory over a ten-day period, a total of 21482 FTP connections.   
Each
connection session is considered as a behavior instance, and we
extract FTP commands in each connection from the dataset.

We computed the frequency distributions of \emph{2}-gram FTP command sequences. Each $2$-gram transition corresponds to a high-level execution feature. We applied the Principal Component Analysis (PCA) technique for dimension reduction, as such a distribution-based profiling produces a large number of features. 
We adopted the X-means clustering approach~\cite{Pelleg:2000:XEK} to cluster all behavior instances in baseline FTP
command sequences, where the center of each of the X-clusters represents a normal program execution context. 
\begin{figure}[h] \vspace{-5pt}
	\centering
	\includegraphics[width=0.7\columnwidth]{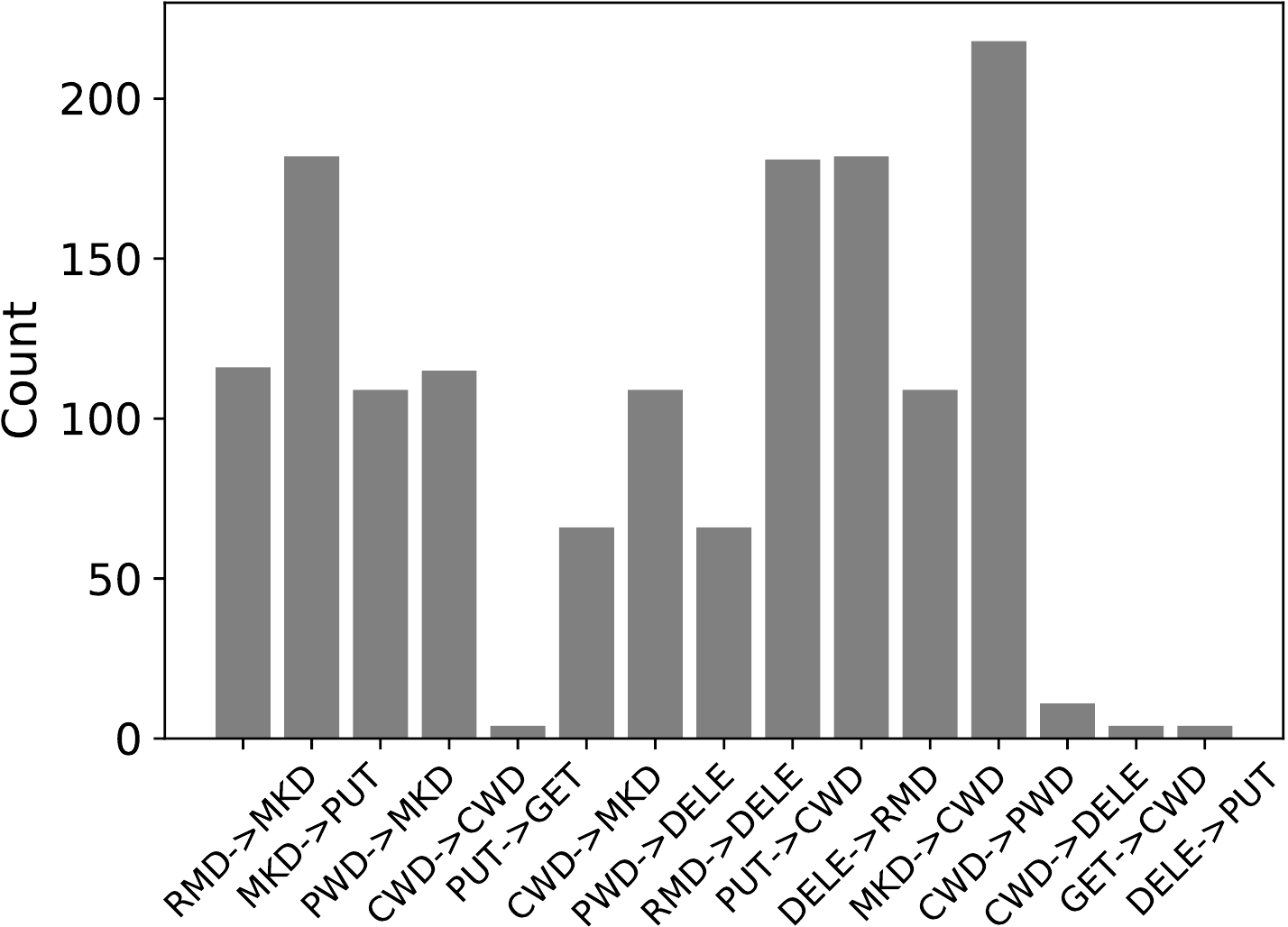} %\vspace{-10pt}
	\caption{For macro-level interaction frequencies, 2-gram distribution of FTP commands within a connection during the ProFTPd DOP attack.}  
	\label{fig:ftpcommand} \vspace{-5pt}
\end{figure}

Fig.~\ref{fig:ftpcommand} shows the macro-level frequency distribution of 2-gram FTP commands within a connection during the ProFTPd DOP attack. Over the course of the attack, it involves more than 1000 client-server FTP commands. In contrast, the average interactions per session in the normal LBNL-FTP-PKT dataset~\cite{LBNL-FTP-PKT} is 41. 

%The ProFTPd DOP attack involves intensive interactions with the server, which triggers more than 180 invocations of the \texttt{sreplace} function. 
\begin{figure}[h] \vspace{-10pt}
	\centering
	\includegraphics[width=0.65\columnwidth]{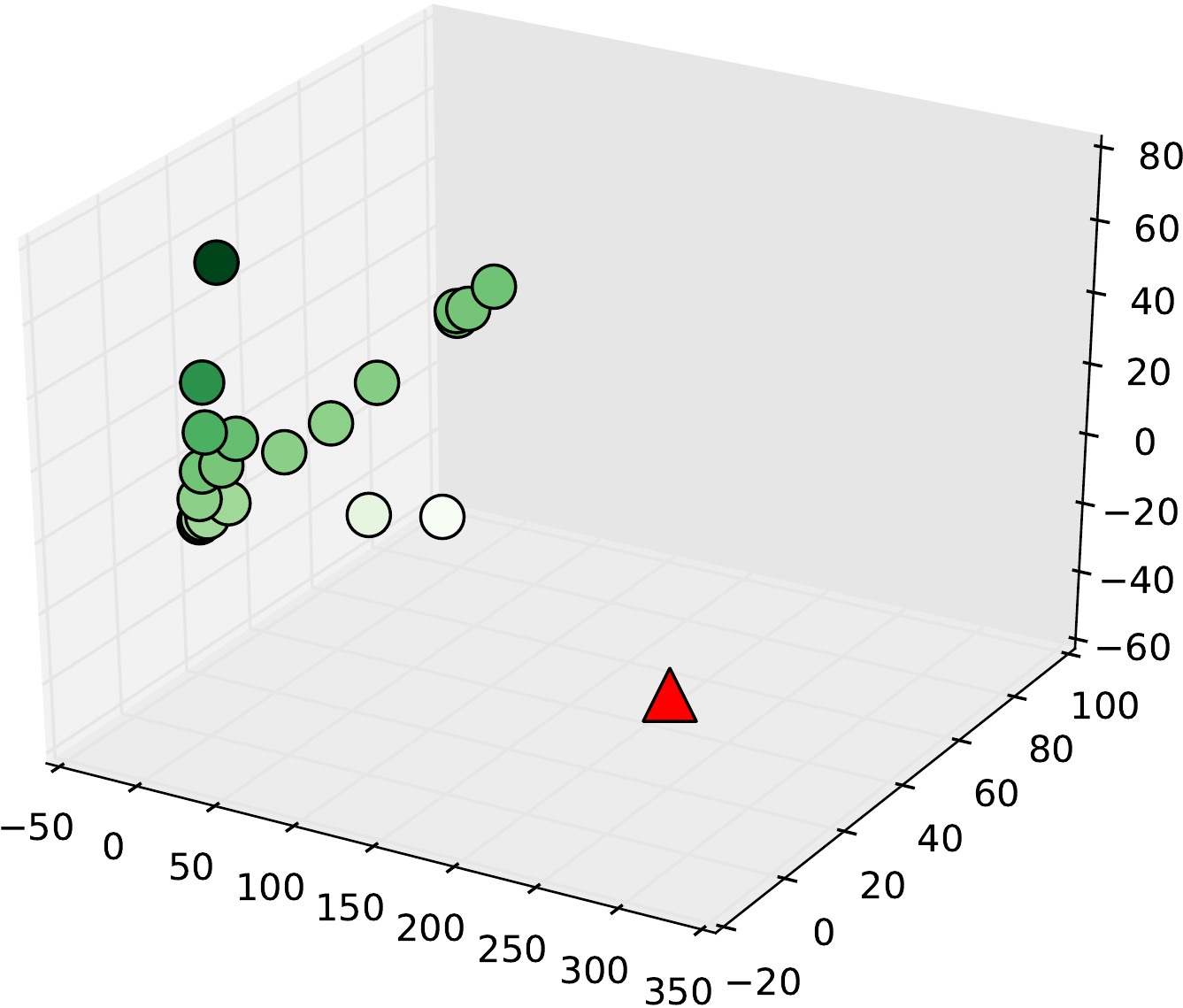} %\vspace{-10pt}
	\caption{For macro-level interaction frequencies, X-clustering for 2-grams of FTP commands with PCA reduction to 3-dimension using LBNL-FTP-PKT dataset~\cite{LBNL-FTP-PKT}. The DOP attack involves an abnormally high number of client-server interactions.}  
	\label{fig:ftpcommandcluster} \vspace{-6pt}
\end{figure}

Fig.~\ref{fig:ftpcommandcluster} illustrates the X-clustering for 2-grams of FTP commands with PCA reduction to 3-dimension. 
The DOP instance (\ie, red triangle) does not belong to any normal clusters (\ie, blue dots). These results suggest that the client-server interactions under the DOP attack drastically differ from the baseline executions.

%CMD --> chdir("/")  DELE: core_dele
%8083296 core_cwd => 80822df _chdir [Loc: mod_core.c:3834 --> mod_core.c:3561] 
%if (strcasecmp(cmd->argv[1], "SITE") == 0)
%return call_module(&site_module, site_dispatch, cmd)

%The normal data is composed of a number of \texttt{sreplace} execution traces collected by PT. 

%Daphne commented below out, no need
%Then, cluster analysis is used to learn distinct normal behaviors from a set of distributions~\cite{IoTDI:Yoon:2017, Shu:2015:USP}. The key difference between our statistical model and~\cite{IoTDI:Yoon:2017, Shu:2015:USP} is that our method embeds ordering information of control-transfers in the model, while solutions in~\cite{IoTDI:Yoon:2017, Shu:2015:USP} neglect the temporal relations among system events, and they work at the system-call level. 

%FTP commands dataset: 21482 FTP connections, 7129 FTP connections with more than 10 seconds per session
%X-clustering for 2-grams of FTP commands per session using the Lawrence Berkeley National Laboratory dataset [10]
%  For a ProFTPd server program under the DOP attack,

\subsubsection{\textbf{Micro-level control-flow frequencies}}
\label{sec:micro-frequency}

% Daphne commented below out -- no need to repeat it again here.
%As mentioned in Section~\ref{sec2:proftpd}, changing the working directory (\ie, \texttt{CWD} command) triggers the invocation of vulnerable \texttt{sreplace} function. In ProFTPd DOP attack, an attacker needs to craft malicious payloads to repeatedly fill up the allocated buffer \texttt{buf} and write bytes beyond the buffer, which exhibits significant frequency anomalies of control-flow transfers in \texttt{sreplace}. 

%[no behavior instance for macro-level interactionsm, only for micro-level control-flow frequencies] When analyzing the macro-level interactions between a FTP server and a client, we define each connection to be a behavior instance, following the approach in~\cite{Shu:2015:USP}. 
%We consider all control-flow transfers in each \texttt{sreplace} invocation as a behavior instance. 

The \texttt{.message} file is the input to vulnerable function \texttt{sreplace}. Its content determines the control-flow behaviors in \texttt{sreplace}, which we measure with PT tracing in this experiment. In the ProFTPd DOP attack, an
attacker crafts \texttt{.message} (\ie, as malicious payloads) to repeatedly fill up
the allocated buffer and write bytes beyond the buffer in \texttt{sreplace},
which exhibits anomalous behaviors of control-flow
transfers. 
We defined all control-flow transfers in each
\texttt{sreplace} invocation as a behavior instance, following the approach in~\cite{Shu:2015:USP}.
Since it is difficult to harvest \texttt{.message} files from
old version FTP servers, in this experiment, we randomly
generated 1000 \texttt{.message} files without triggering the overflow
as the baseline executions. For each \texttt{.message} file, we inserted a character string\footnote{The content of non-specifier character string does not impact on the control-flow behavior in \texttt{sreplace}, which blindly copies non-specifier characters to the output buffer.} with random length less than 100 non-specifier characters and then appended a random specifier (out of 20 different specifiers defined in ProFTPd). We repeated this process with random rounds up to 10 times to generate a \texttt{.message} file.

\begin{figure}[h] \vspace{-10pt}
	\centering
	\includegraphics[width=0.65\columnwidth]{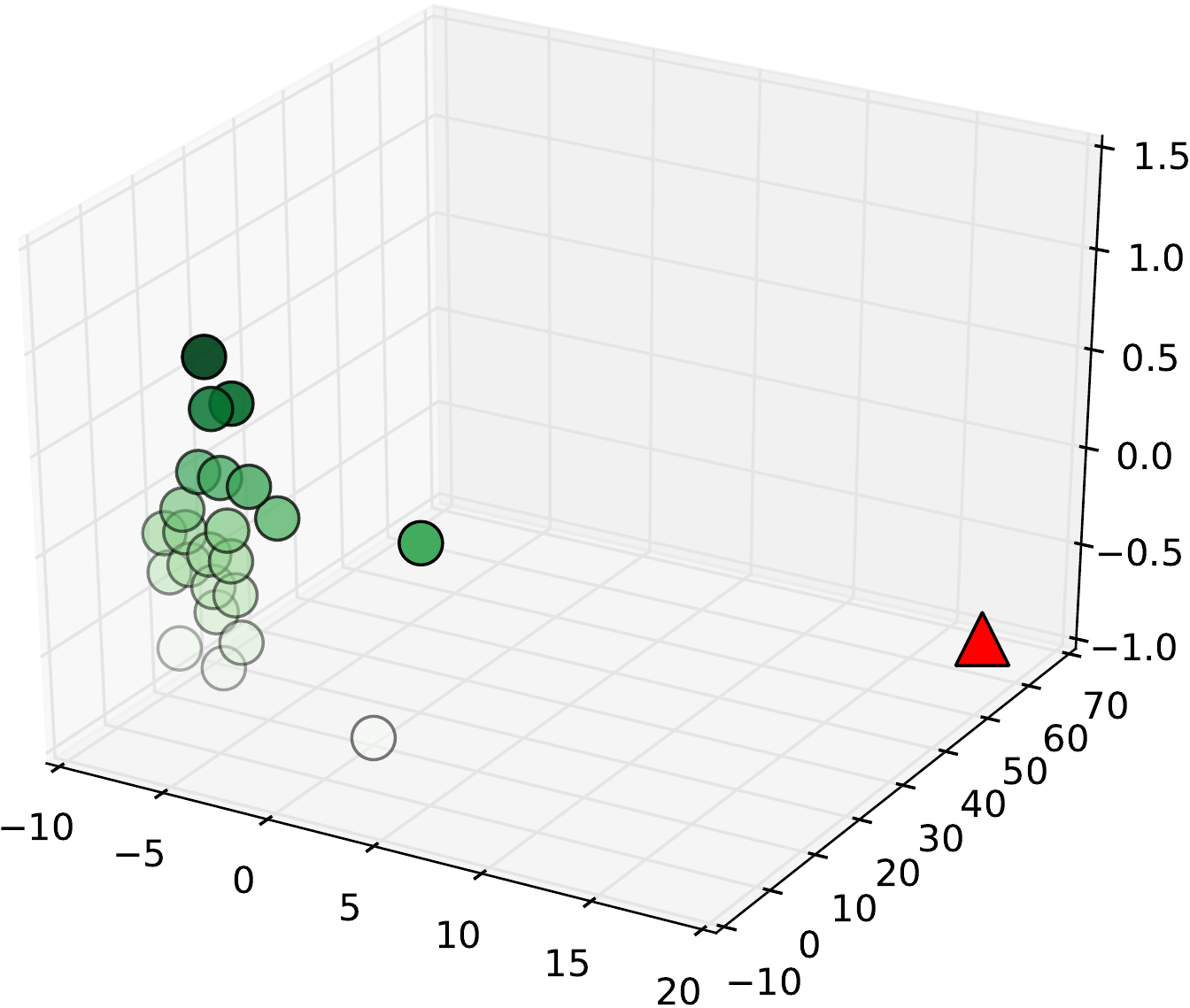} %\vspace{-10pt}
	\caption{For micro-level control-flow frequencies, X-clustering for 2-gram control-transfers with PCA reduction to 3-dimension in \texttt{sreplace}. The DOP attack exhibits a unique pattern of control-flow transfers in comparison to baseline executions.}  
	\label{fig:frequencyanomaly} \vspace{-10pt}
\end{figure}

The feature extraction and dimension reduction procedures are similar to the macro-level analysis described above. After applying PCA, we reduced the original high-dimensional data to 3-dimensional data and then performed the X-clustering. 
Our result comparing the control-flow frequency properties in \texttt{sreplace} is shown in Fig.~\ref{fig:frequencyanomaly}. The baseline dataset is clustered into 23 clusters. Similar to Fig.~\ref{fig:ftpcommandcluster}, the DOP instance is an obvious outlier.
The distance between the DOP instance and any normal cluster is significant, where the average distance is larger than 60-unit length in the 3D space.

%{\bf Long, is this commented out figure worth showing?}

\iffalse
\begin{figure}[h] %\vspace{-10pt}
	\centering
	\includegraphics[width=0.75\columnwidth]{sreplace_controltrans.png} %\vspace{-10pt}
	\caption{2-gram control-flow transition distributions in \texttt{sreplace}}
	\label{fig:controltrans_sreplace} %\vspace{-10pt}
\end{figure}
\fi

%Fig.~\ref{fig:controltrans_sreplace} shows an example of 2-gram control-flow distributions in \texttt{sreplace} compared with the program execution without triggering the overflow. It shows that the DOP attack involves abnormally high number of control-flow transfers in comparison to the baseline execution. 

%Daphne commented below, a bit too vague. safe to comment it out, otherwise, people may ask for more info.
%One can also conduct a similar frequency analysis comparing compatible and  incompatible branch behaviors for data-oriented attacks. 

\subsection{RQ3: Tracing Overhead}\label{sec5:overhead}

%\vspace{-15pt}
\begin{figure}[h]  
	\centering
	\includegraphics[width=0.75\columnwidth]{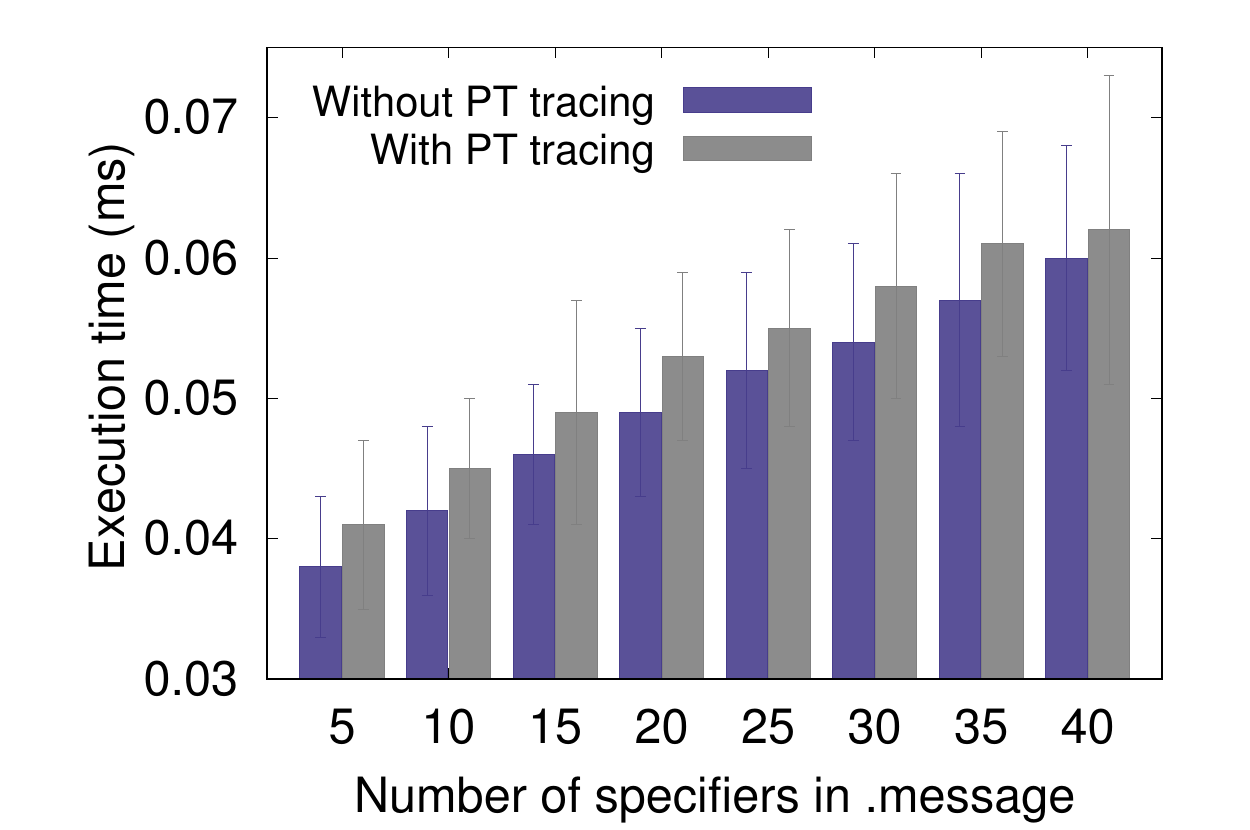} \vspace{-5pt}
	\caption{Tracing overhead of the \texttt{sreplace} function. PT tracing constantly incurs a low overhead ranging from 3.3\% to 7.8\% slowdown in different settings.}  
	\label{chapter5:fig:sreplace:tracing} \vspace{-5pt}
\end{figure}

To measure the PT tracing overhead (\ie, Linux \texttt{Perf-PT}), we instrumented the ProFTPd server program with execution time measurement functions at entry and exit points of the vulnerable function \texttt{sreplace}, and the FTP command handler function \texttt{cmd\_loop}, respectively. We compared the execution time of \texttt{sreplace} as well as the execution time of processing an FTP command with and without PT tracing.
All the results have been averaged over 1000 runs, and the related standard deviations are provided as error bars. 

The control-flow behavior in \texttt{sreplace} depends on the number of specifiers and non-specifier characters in the \texttt{.message} file. In this test, we changed the number of specifiers from 5 to 40 and fixed the length of non-specifier characters to 100 in each \texttt{.message} file. With increasing the number of specifiers, \texttt{sreplace} takes more time to replace these specifiers with dynamic content.
Correspondingly, the number of control-flow transfers involved in \texttt{sreplace} increases from 3181 to 4323. Fig.~\ref{chapter5:fig:sreplace:tracing} shows that the PT tracing overhead is constantly low in different settings. We observed that \texttt{Perf-PT} tracing incurs a low overhead ranging from 3.3\% to 7.8\% slowdown. Processing an FTP command incurs around 2.1\% slowdown (the average execution delay with and without PT tracing is 0.191$ms$ and 0.187$ms$, respectively).

The tracing storage overhead is linearly proportional to the number of control-flow transfers during the program execution. 
To trace control-flow behaviors of \texttt{sreplace}, the size of decoded PT traces ranges from 25.5KB to 34.6KB as the number of specifiers increases from 5 to 40. The low-overhead tracing by PT increases the feasibility for practical deployment of program anomaly detection.

%[ Tracing storage overhead: raw data around [0.20MB]  ]
%[ transfers, 3018, 3181, 3345, 3507, 3670, 3833, 3996, 4159, 4323]
%---------------------------
%Daphne moved below overhead text from section 4, so overhead content stays at the same place

We also experimentally compared the tracing slowdown performance of \texttt{Perf-PT} against three commonly used system-call tracing tools including \texttt{PIN}, \texttt{SystemTap}, and \texttt{Strace}~\cite{Shu:2015:USP}. We measured the elapsed time between the entry and exit points in three utility applications (\ie, \texttt{tcas}, \texttt{replace}, and \texttt{schedule}) from the Software-artifact Infrastructure Repository (SIR) benchmark suite~\cite{sir} on a desktop computer (Ubuntu 16.04, Intel i7-8700 and 32GB of RAM). 

\begin{table}[!htb]
	\begin{center}
		\resizebox{8.8cm}{!}{
			\begin{tabular}{ >{\centering\arraybackslash}m{1.1cm} >{\centering\arraybackslash}m{1.2cm}  >{\centering\arraybackslash}m{1.2cm} >{\centering\arraybackslash}m{1.2cm} >{\centering\arraybackslash}m{1.2cm}>{\centering\arraybackslash}m{1.2cm}  }
				\hline
				\cellcolor{lightgray}{Application} & \cellcolor{lightgray}{Baseline} & \cellcolor{lightgray}{PIN} & \cellcolor{lightgray}{SystemTap} & \cellcolor{lightgray}{Strace} & \cellcolor{lightgray}{Perf-PT} \\
				\hline
				\hline        
				tcas & 0.008$ms$ & 17.59$ms$ & 0.009$ms$ & 0.020$ms$ &  0.009$ms$\\
				\hline
				replace  & 0.038$ms$ & 39.311$ms$ & 0.039$ms$ & 0.079$ms$ &  0.040$ms$ \\
				\hline
				schedule  & 0.047$ms$ & 59.108$ms$ & 0.050$ms$ & 0.086$ms$ & 0.049$ms$ \\
				\hline
			\end{tabular}
		}
	\end{center}	\vspace{-8pt}
	\caption{Average tracing overhead of different tracing tools}
	\label{tab:eva:overhead:pc}	\vspace{-8pt}
\end{table}

The average tracing overhead results out of 1000 runs are shown in Table~\ref{tab:eva:overhead:pc}. The baseline refers to the execution time without tracing. It shows that \texttt{Strace} tracing incurs 113\% slowdown on average, and \texttt{PIN} incurs a rather significant runtime overhead because of the dynamic binary instrumentation. \texttt{SystemTap} and \texttt{Perf-PT} show comparable tracing overhead less than 10\% in our test.
This is because there are a limited number of system-call invocations in these three programs, and thus the binary instrumentation in \texttt{SystemTap} does not introduce too much overhead. However, \texttt{SystemTap} requires a very long instrumentation delay before the program execution. \texttt{Perf-PT} captures all control-flow transfers and thus provides a finer tracing granularity the system-call tracing.

In addition, we measured {\sc Griffin}'s~\cite{Ge:2017:GGC} (a customized PT tracing tool in the Linux kernel) tracing overhead on another Intel PT machine (Intel Core i7-7700 CPU and 32GB of RAM). When tracing \texttt{sreplace} of ProFTPd, we observed 4.33\%$\sim$8.57\% (with a standard deviation ranging from 0.004$ms$ to 0.006$ms$) runtime overhead from {\sc Griffin}, which is consistent to the results using \texttt{Perf-PT}. Since both \texttt{Perf-PT} and {\sc Griffin} follow a trace-all-and-then-filter strategy, their slowdown overhead for tracing the whole program or selectively tracing particular memory regions show very close results. 
We did observe certain PT startup overhead. This one-time PT startup overhead is expected to be amortized across the entire span of the program execution.

%\texttt{Perf-PT} likely adopts this approach as well, based on our observations. The alternative approach is to turn PT on/off to selectively trace certain memory regions, a feature offered by PT. However, this approach is infeasible, due to the repetitive PT startup overheads. 
%but selectively keeps some if a filter is specified.  Selective tracing does impact on the storage overhead, otherwise the PT log size is huge. 

%This result is consistent with the low overhead on \texttt{vsftpd} reported by {\sc Griffin} (Figure 5 in~\cite{Ge:2017:GGC}).\footnote{Low PT tracing overheads indicate low numbers of indirect control transfers.} {\sc Griffin}'s overhead is much lower than \texttt{Perf-PT}'s is likely due to the fact that {\bf Long, please complete this sentence.}

%{\bf Long, Griffin's paper seems show to use 'small cap' as the font for Griffin, that is {\sc Griffin}, as opposed to all caps.}

%\noindent
%{\em Selective PT tracing strategy.} Due to the substantial PT startup overhead, the selective tracing mechanism in {\sc Griffin} follows a trace-all-and-then-filter strategy. The one-time PT startup overhead is amortized across the entire span of the program execution. \texttt{Perf-PT} likely adopts this approach as well, based on our observations. The alternative approach is to turn PT on/off to selectively trace certain memory regions, a feature offered by PT. However, this approach is infeasible, due to the repetitive PT startup overheads. 

%------------------------

\vspace{5pt}
\noindent
{\bf Summary of findings.}
\begin{enumerate}[*]
\item
{\em Branch correlations.} In ProFTPd's vulnerable function \texttt{sreplace} (Listing~\ref{lst:sreplace}), we observed an incompatible
branch behavior, which exposes the DOP's attack footprint. We also developed a branch correlation analysis tool to demonstrate branch correlations with complex predicate expressions are somewhat prevalent in benchmark programs.

%our branch correlation analysis tool identifies 14 coarse-grained correlated conditional branches. *** of them should have subsume relations and *** have mutually exclusive relations, which can be verified at runtime. {\bf Long, would it useful to report these numbers?}
%For the other 8 programs, we found that on average 77\% of the branches have at least one correlated branch. 
These results suggest that the approach of enforcing the {\em branch correlation integrity} (which is a principled defense mechanism), with PT-based branch tracing, could be useful to defend against data-oriented attacks.

\item
{\em Macro-level and micro-level frequencies.} The ProFTPd DOP attack exhibits drastically different frequency distributions.
An attack session triggers the \texttt{sreplace} function more than 180 times. It involves  $>$1000 client-server FTP commands, while the average interactions per session in the normal LBNL-FTP-PKT dataset is 41. The normal and DOP traces (2-gram distributions of PT traces) exhibit strong differences in simple PCA-and-clustering analysis.
 
\item
{\em Runtime and storage overheads.}
Our performance evaluation with \texttt{Perf-PT} and {\sc Griffin} shows that PT's tracing overhead is acceptable, under the trace-all-and-then-filter strategy. The storage overhead is linearly proportional to the number of control-flow transfers during the program execution. With only tracing selective functions or memory regions, the storage overhead is expected to be affordable (in practice, PT log is usually configured as a circular buffer). 

%Daphne commented below out
%Runtime control-flow monitoring without significantly slowing down and with a reasonable storage overhead has the potential to be widely deployed.
%

%{\bf Long, can you add 4 sentences on overhead, time and storage?}

%\item
%We also ported the original ProFTPd DOP attack to a 32bit Ubuntu 16.04 with PT support. We wrote an LLVM-based inter-procedural branch correlation analysis tool to count the numbers of different types of branches with arbitrary predicate expressions.
%{\bf Long, can you open source the code and show URL here?}
\end{enumerate}

% Discssion
\section{Conclusion and Future Research Opportunities}\label{sec7:discussion}

In this SoK work, we systematized the current knowledge on data-oriented exploits and applicable defense mechanisms. 
We experimentally explored the possibility of using low overhead tracing techniques, namely PT, for characterizing data-oriented attacks. We hope that this systematization will stimulate a broader discussion about possible ways to defend against data-oriented attacks. We highlight some interesting future directions in this area. 

%\noindent
{\em Automation of Small Footprint DOP Attacks.}
An interesting research direction is how to minimize the footprints (\ie, side effects) of a DOP attack while achieving the same attack goal. Our experiments in Section~\ref{sec5:detection} showed that DOP alters the correlation or statistical properties of control flows. 
Our empirical study using the FlowStitch benchmarks~\cite{Hu:2015:DOA} revealed that on average 43\% data-oriented gadgets are involved in at least one conditional branch. Gadgets may have different impacts on control-flow behaviors. Attackers may prefer data-oriented gadgets that cause a minimum deviation from normal executions. 
Such a selection process requires automation to be efficient. 
Besides automation, one also needs to define metrics to measure the footprints, \ie, the amount of alteration caused by a DOP execution. Ispoglou~\etal~\cite{Ispoglou:2018:BOP} made the first step towards automating data-oriented programming through a powerful Block Oriented Programming Compiler (BOPC). Searching for gadget chains under specific constraints is a new research direction.

{\em Assessment of Programs' Susceptibility to Data-Oriented Attacks.} 
Such a characterization -- statically or dynamically -- would help one understand the threats that CFI cannot protect against. A promising direction is to quantify the degree of control-flow decisions that are dependent on adversarially controlled data (\eg, user input). Such a characterization also helps prioritize the defense effort, enabling one to address programs with the highest susceptibility first.

%\noindent
{\em Low False Positive PT-based Anomaly Detection.}
DOP attacks exhibit occasional anomalous execution behaviors at runtime, as we have demonstrated in Section~\ref{sec5:detection}. However, to design a successful anomaly detection solution targeting DOP, much more work is needed. Specifically, one needs to show the instruction-level detection does not trigger many false positives in normal executions. Virtually all existing learning-based program anomaly detection demonstrations are at the higher system-call and method-call levels. Reasoning instruction-level PT traces for anomaly detection is challenging.

%\noindent
{\em Deep Learning for Control-Flow Behavior Modeling.} 
Non-control data violations may involve control flows in multiple locations that are far apart. How to detect incompatible control-flow paths, given a relatively long control-flow sequence, is challenging. Exploring deep learning techniques, such as Long Short-Term Memory (LSTM), may be promising, as LSTM keeps track of temporally distant events.

%Deep learning techniques have shown promises in detecting anomalies in different applications~\cite{Deeplearning:survey:2015}. 
%Daphne commented below out. we don't have to cite arxiv papers, plus this paper is short and the work looks flimsy.
%Recent work~\cite{Chen:2018:HeNet} presents a deep learning approach to classify PT generated control-flow traces for malware detection. 
%An important research direction is to apply deep learning algorithms to model program behaviors for anomaly detection.

%\noindent
{\em Selection of Tracing Checkpoints.}
Due to the storage constraint, it is probably impractical to monitor the complete control-flow transfers of a program. Given a limited overhead budget, how to systematically determine strategic checkpoints for tracing (\eg, setting filters to monitor key functions) would be useful in practice.

\bibliographystyle{IEEEtran}
\bibliography{dop-sok-paper}
%\bibliography{main_ref}

% That's all folks!
\end{document}